\documentclass[fleqn,usenatbib]{mnras}

\usepackage{newtxtext,newtxmath}

\usepackage[T1]{fontenc}
\usepackage{ae,aecompl}
\usepackage[utf8]{inputenc}
\newcommand{\angstrom}{\mbox{\normalfont\AA}}

\usepackage{graphicx}	
\usepackage{amsmath}	
\usepackage{amssymb}	

\title[GOGREEN star-forming main sequence]{The GOGREEN survey: The environmental dependence of the star-forming galaxy main sequence at $1.0<z<1.5$}
\author[L. J. Old et al.]{Lyndsay J. Old,$^{1,2}$\thanks{E-mail: lyndsay.old@esa.int}, Michael. L. Balogh$^{3}$, Remco F. J van der Burg$^{4}$, \newauthor Andrea Biviano$^{5,6}$,  Howard K. C. Yee$^{2}$, Irene Pintos-Castro$^{2}$, Kristi Webb$^{3}$, \newauthor Adam Muzzin$^{7}$, Gregory Rudnick$^{8}$, Benedetta Vulcani$^{9}$, Bianca Poggianti$^{9}$, \newauthor Michael Cooper$^{10}$, Dennis Zaritsky$^{11}$,  Pierluigi Cerulo$^{12}$, Gillian Wilson$^{13}$,  \newauthor Jeffrey C. C. Chan$^{13}$,  Chris Lidman$^{14}$,  Sean McGee$^{15}$,  Ricardo Demarco$^{12}$, \newauthor Ben Forrest$^{13}$,  Gabriella De Lucia$^{5}$, David Gilbank$^{16,17}$,  Egidijus Kukstas$^{18}$, \newauthor Ian G. McCarthy$^{18}$, Pascale Jablonka$^{19}$, Julie Nantais$^{20}$,  Allison Noble$^{21,22}$, \newauthor Andrew M. M. Reeves$^{3}$ and Heath Shipley$^{23}$    \\
\noindent Affiliations are listed at the end of the paper
\date{Accepted ??. Received ??; in original form ??}}

\pubyear{2020}

\begin{document}
\label{firstpage}
\pagerange{\pageref{firstpage}--\pageref{lastpage}}
\maketitle

\begin{abstract}

\noindent We present results on the environmental dependence of the star-forming galaxy main sequence in 11 galaxy cluster fields at $1.0 < z < 1.5$ from the Gemini Observations of Galaxies in Rich Early Environments Survey (GOGREEN) survey. We use a homogeneously selected sample of field and cluster galaxies whose membership is derived from dynamical analysis. Using [\ion{O}{II}]-derived star formation rates (SFRs), we find that cluster galaxies have suppressed SFRs at fixed stellar mass in comparison to their field counterparts by a factor of 1.4 $\pm$ 0.1 ($\sim3.3\sigma$) across the stellar mass range: $9.0 <  \log(M_{*} /M_{\odot}) < 11.2$. We also find that this modest suppression in the cluster galaxy star-forming main sequence is mass and redshift dependent: the difference between cluster and field increases towards lower stellar masses and lower redshift. When comparing the distribution of cluster and field galaxy SFRs to the star-forming main sequence, we find an overall shift towards lower SFRs in the cluster population, and note the absence of a tail of high SFR galaxies as seen in the field. Given this observed suppression in the cluster galaxy star-forming main sequence, we explore the implications for several scenarios such as formation time differences between cluster and field galaxies, and environmentally-induced star formation quenching and associated timescales.
\end{abstract}




\section{Introduction}
Measurements of the galaxy stellar mass function and cosmic star formation rate (SFR) as a function of redshift have demonstrated that the global star formation activity of galaxies peaked at $z\sim2$, declining until the present day (e.g. \citealt{Madau_2014} and references therein). This evolution is also seen as a decrease in the specific SFR (sSFR) of galaxies with time since $z\sim 2$ (e.g. \citealt{Whitaker_2012} and others), and is  characterised as evolution in the correlation between SFR and stellar mass, referred to as the star-forming main sequence \citep{Noeske_2007}.  However, comparing the evolution of the stellar mass functions for star-forming and quiescent galaxies separately \citep{Peng_2010,Muzzin_2013b} shows that a stellar mass-dependent "quenching" of star formation must also be taking place.  This quenching refers to a comparatively rapid terminal cessation of star formation that leads to the gradual build-up of the passively-evolving galaxy population.\\
\indent There is also evidence that the evolution of galaxies depends on their environment - whether this means local density, or their location as a satellite galaxy within a more massive host dark matter halo.  At $z<1$, galaxies in denser environments such as galaxy groups and galaxy clusters universally have lower fractions of star-forming galaxies than the field \citep[e.g.][]{Kauffmann_2004, Balogh_2004a, Poggianti_2006, Cooper_2006, Cooper_2007, Kimm_2009, vonderlinden_2010, Peng_2010, Muzzin_2012, Mok_2013, Davies_2016, Guglielmo_2019, Pintos_Castro_2019}. 

One explanation for this trend is that galaxies in groups and clusters are subject to additional processes that enhance the quenching rate \citep[e.g.][]{Balogh_2004a, Peng_2010, Wetzel_2013}. If this is the case, there should exist a transition galaxy population in groups and clusters that have low but non-zero SFRs.  The bimodality in the galaxy colour and SFR distribution \citep[e.g.][]{Strateva_2001, Baldry_2004, Balogh_2004b, Cassata_2008, Wetzel_2012, Taylor_2015} suggests that these transition galaxies are rare, implying a rapid transformation from the star-forming to quiescent population. Identifying these transition galaxies from their lower-than-average SFRs requires large, carefully-selected samples over a wide stellar mass range, and results to-date are mixed. Several studies have claimed little to no trend in the star-forming main sequence with environment \citep[e.g.][]{Peng_2010, Wijesinghe_2012, Muzzin_2012, Wetzel_2012,Koyama_2013}; others find a modest trend in the sense that star-forming galaxies in denser environments have lower star formation rates at fixed stellar mass than that of their counterparts in the field  \citep[e.g.][]{Vulcani_2010, vonderlinden_2010, Popesso_2011, Patel_2011, Haines_2013, Paccagnella_2016, Rodriguez_2017, Wang_2018}. 

In order to reconcile the modest, at best, differences between the SFRs of galaxies in groups, clusters and the field at low redshift with the fact that clusters host a much larger fraction of quenched galaxies, \citet{Wetzel_2013} introduced a two-parameter model to describe the suppression of star formation for satellites in massive haloes.  In this `delayed-then-rapid’ quenching scenario, as a satellite galaxy infalls into a cluster, there is a period of time within which a galaxy's SFR follows that of typical field galaxy evolution. After this `delay-time', a galaxy experiences a swift truncation in its SFR. 

Using a galaxy group/cluster catalogue (based on \citealt{Yang_2005}) from the Sloan Digital Sky Survey Data Release 7 at $0.04<z<0.06$ \citep{York_2000, Abazajian_2009}, together with a high-resolution, cosmological $N$-body simulation to track satellite orbits, \citet{Wetzel_2013} empirically fits a delayed-then-rapid quenching scenario where galaxy SFRs are unaffected for 2-4~Gyr following infall, after which star formation quenches rapidly. The long delay time is somewhat puzzling, given the shorter dynamical times associated with galaxy orbits. However, some authors \citep[e.g.][]{Taranu_2014,Oman_2016,Muzzin_2014} find good agreement with models where quenching begins only when galaxies first pass within a small radius near the cluster core or where environmental quenching is driven by halting gas accretion \citep[e.g.][]{Fillingham_2015}.

A promising approach to better understand these results is to look at higher-redshift clusters and groups.  Because the dynamical time for a virialized system is shorter at higher redshift, and independent of halo mass \citep{Tinker_2010,Tinker_2013,McGee_2014}, we might hope to determine if these quenching timescales are associated with orbital parameters. An alternative might be that they are determined by properties of the galaxy itself (e.g.~gas content and star formation rate); these generally evolve at a different rate from the dynamical time, allowing us to break the degeneracy \citep{McGee_2014}. Observations probing groups and clusters at intermediate redshifts are generally consistent with a total quenching time (i.e.~the time between infall and cessation of star formation) that evolves approximately like the dynamical time \citep[e.g.][]{Tinker_2010, Mok_2014, Balogh_2016, Foltz_2018}. However, there are indications that the {\it delay} time at $z\sim1$ is shorter than would be expected from a simple scaling with dynamical time from $z\sim0$ \citep{McGee_2014}. These authors suggest that the high SFRs associated with galaxies coupled with mass-loaded winds at $z\sim1$, means that they will exhaust their gas supply on a timescale that is shorter than the dynamical time and hence quench before any orbit-related process like ram-pressure stripping can be effective. This phenomenon, called "overconsumption", also proves to be a good match to the stellar mass dependence of the observed quenched galaxy fraction at $z\sim1$ \citep{Balogh_2016, Kawinwanichakij_2017, Fossati_2017, Chartab_2019}.

It is therefore interesting to extend this work to even higher redshift, $z>1$, where galaxies are significantly younger, have higher SFRs and lower depletion timescales (e.g.~\citealt{Tacconi_2013,Tacconi_2018}).  While there are pioneering studies of galaxy clusters at higher redshifts, (e.g.~$z > 1$; \citealt{Strazzullo_2006, Gobat_2008, Snyder_2012, Lotz_2013, Zeimann_2013, Martini_2013, Nantais_2013b, Newman_2014, Stanford_2014, Nanayakkara_2016, Nantais_2017, Strazzullo_2019}), deep spectroscopic cluster studies of galaxies of all types for a range of halo masses do not currently exist, and the typical properties of cluster galaxies at this epoch are still unknown.

The goal of this study is to measure the difference in the star-forming galaxy main sequence between cluster and field galaxies with a deep spectroscopic sample of homogeneously targeted galaxies above $z >1$. For this purpose, we use the recently completed Gemini Observations of Galaxies in Rich Early ENvironments (GOGREEN) survey \citep{Balogh_2017}. In Section~\ref{sec:gogreen_sample}, we describe the key survey details including the cluster sample and some aspects of the data reduction. In Section~\ref{sec:Results}, we present results on the environmental dependence of the star-forming main sequence and the corresponding discussion in Section~\ref{sec:discussion}. We conclude our findings in Section~\ref{sec:conclusions}. Throughout the paper we adopt a flat Lambda Cold dark matter ($\Lambda$CDM) cosmology with $\Omega_{\rm m}=0.3$, and a Hubble constant of $H_{\rm 0} = 70$ $\rm km~s^{-1}$ $\rm Mpc^{-1}$. We also assume a Chabrier Initial Mass Function (IMF; \citealt{Chabrier_2003}).

\section{The GOGREEN Survey}
\label{sec:gogreen_sample}
The GOGREEN survey is based on a Gemini Large and Long Program using the GMOS instruments \citep{GMOS_1998, Hook_2004} on Gemini North and South telescopes to obtain unbiased multi-object spectroscopy of galaxies of all types down to stellar masses $M_{*} \sim 10^{10.3}~\rm{M_{\odot}}$\footnote{A list of all GOGREEN papers can be found on the following webpage: \href{http://gogreensurvey.ca/data-releases/publicationspress/}{http://gogreensurvey.ca/data-releases/publicationspress/}.} . The survey targets 21 systems spanning a wide range in halo mass at $1.0 < z < 1.5$. In addition to deep spectroscopy, the GOGREEN survey has obtained over one hundred hours of $UBVRIzYJK$ and IRAC 3.6$\mu$m imaging for the majority of the target fields, providing photometric products including stellar masses, accurate rest-frame $UVJ$ colour measurements and characterisation of spectroscopic completeness. Crucially, the survey is designed to produce a field sample of comparable size to that of the targeted groups and clusters, selected under the same conditions. 

One of the key science goals of GOGREEN is to probe environmental quenching and the growth of the stellar mass function at this early epoch. The survey is also designed to examine the stellar populations and dynamics of galaxies in clusters that span a wide range in halo mass at $1.0 <z< 1.5$ when the Universe was $<6$ Gyr old. Candidate groups and clusters are selected in three approximate bins of mass: groups ($M < 10^{14}~\rm{M_{\odot}}$), typical clusters ($10^{14} <M/\rm{M_{\odot}}< 5\times10^{14}$) and very massive clusters ($M > 5\times10^{14}~\rm{M_{\odot}}$). Within GOGREEN, we target nine group mass candidates in the COSMOS \citep{Finoguenov_2007,George_2011} and Subaru XMM Deep Survey \citep[SXDS,][]{Finoguenov_2010} fields, nine typical clusters from the {\it Spitzer} Adaptation of the Red-sequence Cluster Survey \citep[SpARCS,][]{Wilson_2009, Muzzin_2009, Demarco_2010b}, five of which have
extensive GMOS spectroscopic follow-up from the Gemini Cluster Astrophysics Spectroscopic Survey \citep[GCLASS,][]{Muzzin_2012}, and we target three very massive clusters from the South Pole Telescope (SPT) survey \citep[]{Brodwin_2010, Foley_2011, Stalder_2013}.
In this study, we focus on eleven of the GOGREEN clusters, the properties of which are described in Section~\ref{sec:cluster_membership} and summarised in Table~\ref{tab:cluster_sample_table}. 

\subsection{Spectroscopic data}
In this Section, we summarise the relevant spectroscopic observations, spectroscopic targeting and reduction for the data used in this study. Full details are given in the survey description paper \citep{Balogh_2017} and a forthcoming data release paper (Balogh et al., in preparation). Spectroscopic targets were selected using magnitude and colour cuts from combined deep GMOS $z'$-band imaging and {\it Spitzer} IRAC 3.6$\mu$m photometry. The GMOS $z'$-band imaging was obtained as part of the GOGREEN program, and $5\sigma$ depths range from 24.75 -- 25.70 for the eleven GOGREEN clusters used in this study. More details regarding the integration time, condition, and depths for each individual field can be found in \citet{Balogh_2017}.

We focus on 11 of the 12 GOGREEN cluster fields, omitting the twelth GOGREEN cluster, SpARCS1033, from this analysis, as the K-band observations are not yet complete. We complemented our own deep 3.6$\mu$m {\it Spitzer} IRAC imaging with publicly-available imaging ($5\sigma$ AB depth of at least 2$\mu$Jy or AB=23.1) from SERVS \citep{Mauduit_2012}, S-COSMOS \citep{Sanders_2007}, and SpUDS \citep{Galametz_2013}, and additional CO programs (PI=Menanteau, PID=70149, PI: Brodwin, PIDs=60099,
70053). Targeted galaxies are selected to have total magnitudes of [3.6] < 22.5 and $z'$ < 24.25, avoiding low-redshift ($z<1$) contamination by imposing a colour cut determined using the colour-magnitude distribution of galaxies with high-quality photometric redshifts in UltraVISTA \citep{Muzzin_2013b}. Spectroscopic slit masks were designed in order to obtain high numbers of bright and faint galaxies while also ensuring reasonable completeness in the cluster core. For further details regarding mask design, and for technical details regarding data reduction, we refer the reader to Section 2.4.3 and Section 3 in \citet{Balogh_2017}, respectively.

Most observations were obtained with the upgraded Hamamatsu detectors on GMOS-N and GMOS-S \citep{Gimeno_2016}, though some of the earliest data taken on Gemini-N used the older EEV deep depletion detectors.  All fields were observed with the R150 grating to maximise the wavelength coverage (observed wavelength range of the spectra is $\sim 5500-10500$\AA) on the detector and ensure high redshift completeness. . Nod-and-shuffle mode was used to ensure good sky subtraction at red wavelengths, and to maximise slit density in the cluster cores.   All observations were obtained with the detector binned $2\times2$, delivering a dispersion of 3.8$\angstrom$/pix for the Hamamamatsu detectors and 3.5$\angstrom$/pix for the EEV detector. With slit widths of 1", the resulting spectral resolving power is $R \sim 460$ (with spectral resolution $\sim20$\AA/pix).  A relative flux calibration is applied to the spectra based on standard star observations taken once per semester. Absolute flux calibration is described in Section~\ref{sec-photo}. Telluric absorption is corrected using {\sc molecfit} \citep{Smette_2015, Kausch_2015}, and redshifts are computed via cross-correlation with a variety of templates, using {\sc marz} \citep{Hinton_2016}.

GOGREEN deliberately builds upon the previous lower-redshift galaxy cluster survey GCLASS \citep{Wilson_2009, Muzzin_2009,Demarco_2010b}, for which data was taken in a very similar manner (similar exposure times), and we incorporate GMOS spectroscopy for five of the $z > 1$ clusters from that survey in our analysis.

\subsection{Photometric data}\label{sec-photo}
We use $K$-band selected photometric catalogues derived from deep, multi-band imaging (van der Burg et al. in preparation). Stellar masses are derived from SED fitting to multiwavelength photometry, using {\sc FAST} \citep{Kriek_2009, Kriek_2018} and stellar population synthesis models from \citet{Bruzual_2003}. A Chabrier IMF \citep{Chabrier_2003}, solar metallicity, and the dust law from \citet{Calzetti_2000} are assumed.  The star formation history is parameterised  as $SFR \propto e^{-t/\tau}$, where $\tau$ ranges between 10 Myr and 10 Gyr, and the age is left as a free parameter. We note that stellar masses derived from non-parametric star formation histories (SFH) have been found to be typically $\sim 0.2$~dex higher than those derived in this work (\citealt{Leja_2019a,Leja_2019b}, Webb et al. in preparation). 

\begin{table*}
	\centering
	\caption{This table summarises the key properties of the eleven clusters used in this study that are targeted by the GOGREEN survey. $z_{\rm c}$ is the initial cluster mean redshift estimate, taken from \citet{Balogh_2017}, with the exception of SpARCS0219, for which the $z_{\rm c}$ estimates have been obtained using the peak of the $z$ distribution. $N_{\rm tot}$ is the number of objects with good quality redshifts in the cluster field from either GOGREEN or the literature, avoiding double entries, and $N_{\rm mem}$ is the number of galaxies that are considered as members by at least one of the two C.L.U.M.P.S and Clean algorithms. $\sigma_{\rm los}$ is the rest-frame line-of-sight velocity dispersion of the cluster obtained by using the galaxy membership probabilities as weights. Finally, $r_{\rm 200c}$ is the overdensity radius of the cluster (where the overdensity 200 times critical at the cluster redshift), estimated from $\sigma_{\rm los}$ using Equation B3 in
	\citet{Mamon_2013}.}
	\label{tab:cluster_sample_table}
	\begin{tabular}{cllccccc} 
		\hline
		Cluster name & $\rm RA_{\rm \;J2000}^{\rm \;BCG}$ & $\rm DEC_{\rm \;J2000}^{\rm \;BCG}$ & $z_{\rm c}$ & $N_{\rm tot}$ & $N_{\rm mem}$ & $\sigma_{\rm los}$ $[\rm{km/s}]$ & $r_{\rm 200c}$ $[\rm{Mpc}]$\\
		\hline
		SPT--CLJ0205--5829 & 02:05:48.19 & --58:28:49.0 & 1.320 & 70  & 28 & 678 $\pm$ 57 & 0.76 $\pm$ 0.09\\
		SPT--CLJ0546--5345 & 05:46:33.67 & --53:45:40.6 & 1.067 & 103 & 67 &  977 $\pm$ 68 & 1.17 $\pm$ 0.09 \\
		SPT--CLJ2106--5844 & 21:06:04.59 & --58:44:27.9 & 1.132 & 81 & 50 & 1055 $\pm$ 83 & 1.23 $\pm$ 0.10\\
		\hline
		SpARCS0035--4312 & 00:35:49.68 & --43:12:23.8 & 1.335& 129 & 29 & 840 $\pm$ 52 & 0.93 $\pm$ 0.07\\
        SpARCS0219--0531 & 02:19:43.56 & --05:31:29.6 & 1.325 & 56 & 9 & 810 $\pm$ 77 & 0.79 $\pm$ 0.12\\
        SpARCS0335--2929 & 03:35:03.56 & --29:28:55.8 & 1.368 & 133 & 27 & 542 $\pm$ 33 & 0.67 $\pm$ 0.08\\
        SpARCS1034+5818 & 10:34:49.47 & +58:18:33.1 & 1.386 & 40 & 11 & 250 $\pm$ 28 & 0.24 $\pm$ 0.03\\
        SpARCS1051+5818 & 10:51:11.23 & +58:18:02.7 & 1.035 & 185 & 42 & 689 $\pm$ 36 & 0.88 $\pm$ 0.07\\
        SpARCS1616+5545 & 16:16:41.32 & +55:45:12.4 & 1.156 & 214 & 60 & 782 $\pm$ 39 & 0.92 $\pm$ 0.06\\
        SpARCS1634+4021 & 16:34:37.00 & +40:21:49.3 & 1.177 & 190 & 69 & 715 $\pm$ 37 & 0.85 $\pm$ 0.06\\
        SpARCS1638+4038 & 16:38:51.64 & +40:38:42.9 & 1.196 & 174 & 56 & 564 $\pm$ 30 & 0.70 $\pm$ 0.06 \\
		\hline
	\end{tabular}
\end{table*}

To obtain an absolute calibration for the GOGREEN spectra, we use the appropriate $I$-band photometry. We first interpolate the filter response curve, $R$, to match the spectral wavelength distribution using cubic-spline interpolation. We then integrate over the interpolated filter response curve multiplied by the spectral flux, $f_{\rm \lambda}$ to give the total spectral flux:
\begin{equation}
f_{\rm tot} = \int_{\lambda_{\rm min}}^{\lambda_{\rm max}}(R f_{\lambda} \lambda d\lambda)/\int_{\lambda_{\rm min}}^{\lambda_{\rm max}}(R \lambda d\lambda) \; .
\end{equation}
After converting this spectral flux in wavelengths to frequency, we then multiply the entire spectra by the ratio of the spectral flux and the total $I$-band flux derived from the photometry to obtain flux-calibrated spectra corrected for slit losses.

\subsection{Cluster membership}
\label{sec:cluster_membership}
We use the dynamical properties of the galaxies to select those that are cluster members, employing two methodologies. One approach, referred to as the Clean algorithm \citep{Mamon_2013}, uses an estimate of the cluster line-of-sight velocity dispersion,  $\sigma_{\rm los}$, to predict the cluster mass from a scaling relation. The other algorithm, referred to as C.L.U.M.P.S (Munari et al. in preparation) is based on the Shifting Gapper (SG) method of \citet{Fadda_1996}. In this paper, cluster members are defined as those that are identified by either the Clean or C.L.U.M.P.S algorithm. We refer the reader to Section~\ref{sec:appendix_cluster_membership} in the appendix for more details regarding these membership algorithms.

In Table~\ref{tab:cluster_sample_table}, we list the cluster mean redshift $\bar{z}$, the number of objects with good quality redshifts in the cluster field from either GOGREEN or the literature, $N_{\rm tot}$, and the number of member galaxies that are considered members by at least one of the two C.L.U.M.P.S and Clean algorithms, $N_{\rm mem}$. We also include the rest frame line-of-sight velocity dispersion, $\sigma_{\rm los}$, derived from this dynamical membership procedure, and an estimate of $r_{\rm 200c}$ for each cluster in Table~\ref{tab:cluster_sample_table}. We note that our spectroscopic targeting extends beyond $r_{\rm 200c}$, but does not equally sample the area beyond this radius. Throughout this paper, the centre of the cluster is taken as the location of the BCG when available. In this work, the BCG is defined as the most massive galaxy with photometric redshift consistent with the cluster mean redshift, and projected within 500 kpc from the main galaxy over-density (for more details, we refer the reader to van der Burg et al. in preparation). We note that excluding potential Active Galactic Nuclei (AGN) candidates (identified via several diagnostics), produces no qualitative change in our results. From works such as \citet{Martini_2013}, we expect a consistent AGN fraction in our cluster galaxy sample with respect to the field at $1.0 < z < 1.5$. We also note that the expected fraction in clusters is small (e.g., $< 0.1$), and that many of these AGN are likely to be in early-type galaxies (e.g., \citealt{Kauffmann_2003}).

\subsection{[OII] detection}
\label{sec:OII_measurements}
\begin{figure}
	\includegraphics[width=1.0\columnwidth]{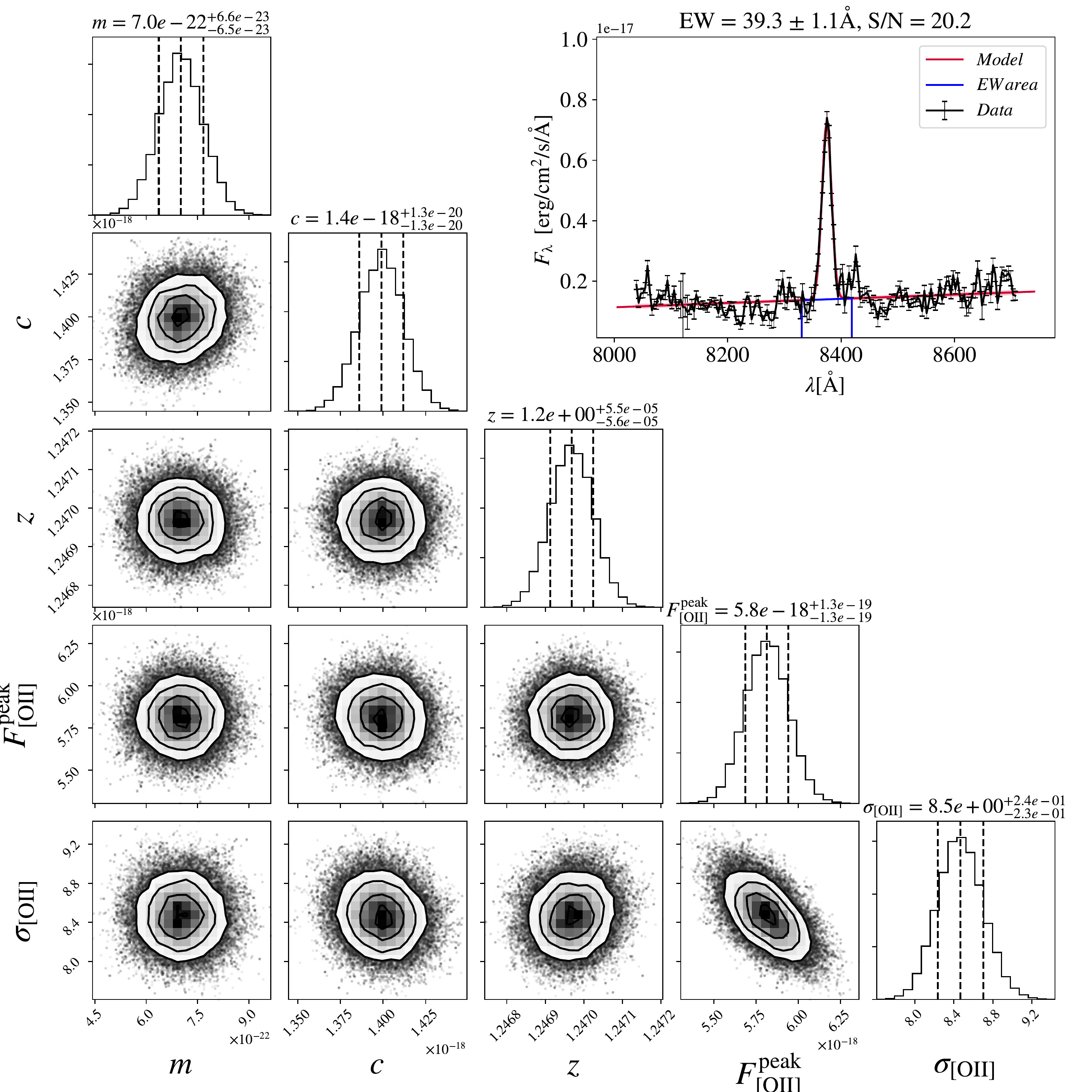}
    \caption{Example of the marginalized probability distributions produced
by our MCMC analysis of [\ion{O}{II}] emission for the five parameter model of a composite of a linear continuum and a Gaussian emission line. The parameters are the slope, ($m$), the intercept ($c$), the galaxy redshift ($z$), the peak flux of the emission ($F^{\rm peak}_{[\ion{O}{II}]}$), and the width of the [\ion{O}{II}] emission line ($\sigma_{[\ion{O}{II}]}$).}
    \label{fig:OII_MCMC_corner_posteriors_and_fit_SpARCS1634_111000817}
\end{figure}

\begin{figure}
	\includegraphics[width=1.0\columnwidth]{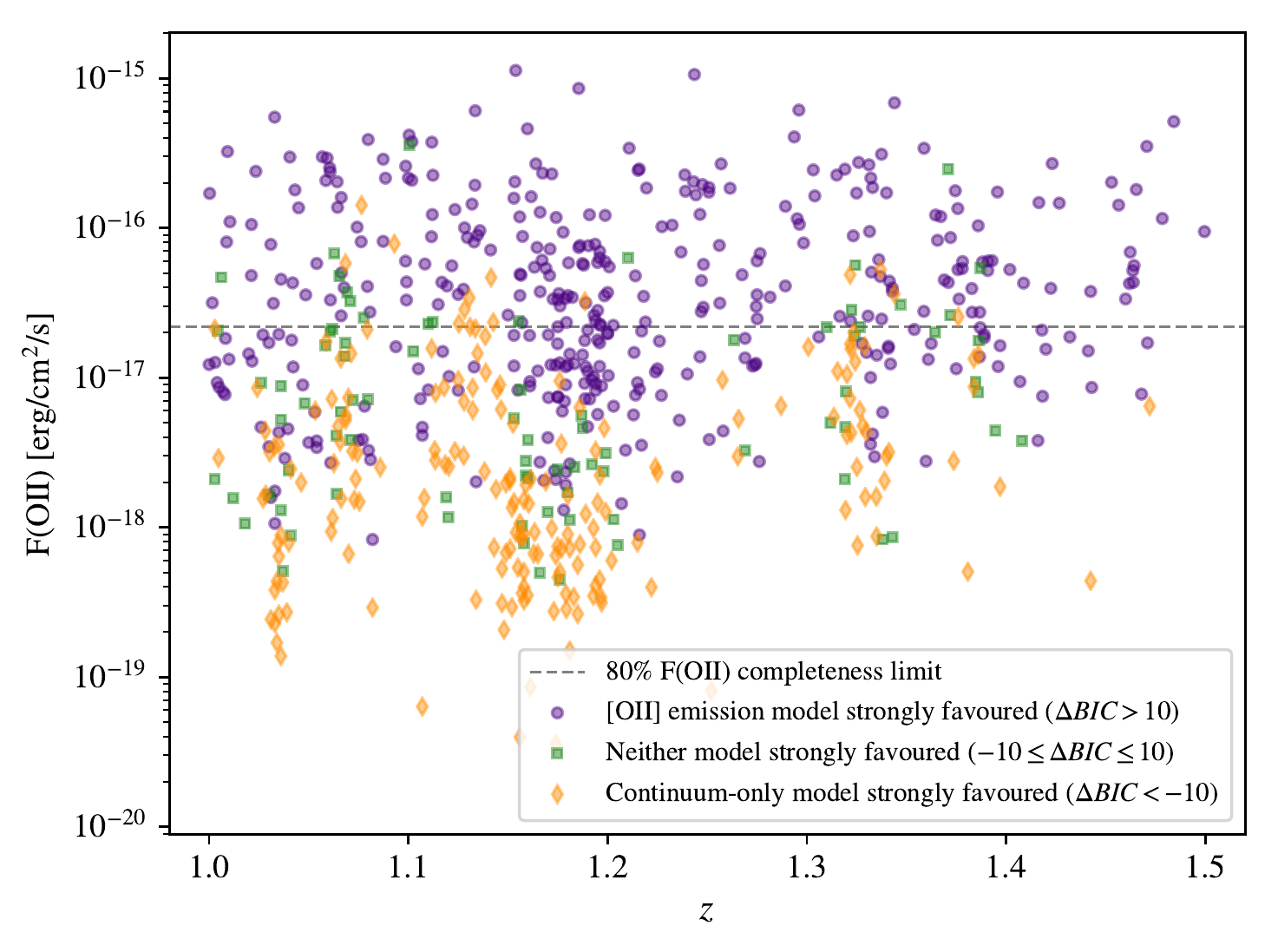}
    \caption{Galaxy redshift and $F([\ion{O}{II}])$ where points are colour-coded by the $\Delta {\rm BIC}$ criteria described in Section~\ref{sec:OII_measurements}. Galaxies where the model including both a continuum and an [\ion{O}{II}] emission line are strongly favoured over a model with just a continuum are represented as purple circles. Galaxies where neither model is strongly favoured are represented as green squares, and galaxies where the continuum-only model is strongly favoured are represented as dark orange diamonds. The $80\%$ flux completeness limit, $F([\ion{O}{II}])=2.2\times10^{-17}~{\rm erg}~{\rm cm}^{-2}~{\rm s}^{-1}$, is shown as a dashed gray line. Galaxies above this flux limit are selected as star-forming, and are used for the subsequent analysis.}
    \label{fig:z_vs_log_F_OII}
\end{figure}

We take a Bayesian model selection approach in detecting [O\,{\sevensize II}] emission in the GOGREEN galaxy spectra. Separately, we fit two models to the data, with the first assuming a linear continuum and the second assuming a composite of a linear continuum and a Gaussian emission line according to:
\begin{equation}
    f_{\rm model}(\lambda)=m\;\lambda+c+\mathcal{N}(\lambda_{[\ion{O}{II}]}(1+z)|F^{\rm peak}_{[\ion{O}{II}]}, \sigma_{[\ion{O}{II}]}) \; .
\label{eq:cont_plus_gaussian_fit_eq}
\end{equation}
Both models depend on the following three parameters: $m$ and $c$ are the slope and intercept of the continuum, and $z$ is the galaxy redshift. The second model also depends on two parameters describing the Gaussian-shaped emission: $F^{\rm peak}_{[\ion{O}{II}]}$ is the peak flux of the emission and $\sigma_{[\ion{O}{II}]}$ is the width of the [\ion{O}{II}] emission line. We assume that our uncertainties follow a normal distribution, and therefore our likelihood function for both models is
\begin{equation}
\ln \mathcal{L}=-\frac{1}{2}\sum^{n}_{i=1} \left( \frac{f_{i}-f_{\rm model}}{\sigma_{i}}\right)  ^{2}=-\frac{1}{2}\chi^{2} \; .
\label{eq:likelihood}
\end{equation}
We employ {\sc emcee}, an affine-invariant ensemble sampler for MCMC, to efficiently explore our parameter space and determine the posteriors of the model parameters \citep[]{Foreman_2013}. We utilize 50 walkers and perform 800 iterations per model per spectrum (including a `burn-in' of 300 iterations) assuming flat priors for model parameters. We fit a section of the spectra within the wavelength range of $\pm 150$\AA~from the predicted location of the [\ion{O}{II}] emission line given the galaxy redshift. An example of the marginalized probability distributions produced by our MCMC analysis for the five-parameter model is demonstrated in Figure~\ref{fig:OII_MCMC_corner_posteriors_and_fit_SpARCS1634_111000817}.
We use Bayesian Information Criterion (BIC) as a criterion for model selection among the two models described above, one which assumes an emission line is present at the expected rest-frame wavelength of [\ion{O}{II}], and one which assumes there is no emission line present at the rest-frame wavelength of [\ion{O}{II}].  BIC is  based on the likelihood function, and takes into account the number of parameters in a model so as to penalise models with more parameters to avoid a possible increase in the likelihood solely by increasing the number of parameters \citep[]{Schwarz_1978}. The form for calculating the BIC is: 
\begin{equation}
{\rm BIC} = k \ln(n)- 2 \ln(\mathcal{\hat{L}}) \; ,
\label{eq:BIC}
\end{equation}
where $n$ is the number of data points, $k$ is the number of parameters estimated by the model, and $\mathcal{\hat{L}}$ is the maximum value of the likelihood function of the model. The model with the lowest BIC is preferred, with the degree of model favourability adopted by the classification of \citet{kass_1995}, which takes into account the $\Delta{\rm BIC}$. We take a conservative approach, adopting a criterion of $\Delta{\rm BIC}  > 10$ to ensure that the model with an emission line is favoured only with `very strong' evidence against higher BIC. Spectra where  $\Delta{\rm BIC} < 10$ are comprised of cases where evidence for supporting a model where an emission line is weak or where evidence for a model without an emission line is preferred. In Figure~\ref{fig:z_vs_log_F_OII} we show the galaxy redshift and $F([\ion{O}{II}])$ for all galaxies (regardless of membership) where points are colour-coded by the $\Delta{\rm BIC}$ criteria. We note that adjusting these criteria from `very strong' evidence to `strong' evidence ($\Delta{\rm BIC}  > 6$) has little effect on the number distributions of objects in these categories.

To deduce the minimum [\ion{O}{II}]-derived flux, $F([\ion{O}{II}])$, with which to securely select [\ion{O}{II}] detections, we bin galaxies by flux and calculate the percentage of galaxies within each flux bin where the emission line model is very strongly favoured according to the BIC criterion. We then fit this relation between flux and ratio of model criterion with a fourth degree polynomial and define the minimum $F([\ion{O}{II}])$ as the flux at which this percentage reaches 80$\%$. With this conservative limit of $F([\ion{O}{II}])=2.2\times10^{-17}~{\rm erg}~{\rm cm}^{-2}~{\rm s}^{-1}$, there are 262 total [\ion{O}{II}] detections, including 100 cluster galaxies, 162 field galaxies.

\subsection{[OII]-derived star formation rates}
\label{sec:OII_SFRs}
The [\ion{O}{II}]-derived star formation rates for the GOGREEN galaxies are calculated from the measured [\ion{O}{II}] luminosities according to the relation from \citet{Gilbank2010}, where:
\begin{equation}
{\rm SFR}_{0}/\left({\rm M}_{\odot}~{\rm yr}^{-1}\right)=L([\ion{O}{II}])/(3.80\times10^{40}~{\rm erg}~{\rm s}^{-1}) \; .
\label{eq:SFR0}
\end{equation}
We use the empirical correction derived from H$\alpha$ to correct this nominal star formation rate (${\rm SFR}_{0}$) for the metallicity and dust dependence of [\ion{O}{II}] luminosity on SFR as a function of stellar mass, such that corrected SFR is given by:
\begin{equation}
{\rm SFR} = \frac{{\rm SFR}_{0}}{a \; {\rm tanh}[(x-b)/c] + d} \; ,
\label{eq:SFRempcorr}
\end{equation}
where $x=\log(M_{*}/{\rm M}_{\odot})$, $a=-1.424$, $b=9.827$, $c=0.572$ and $d=1.700$. For more details regarding the empirically-derived correction, we refer to \citet{Gilbank2010}. The SFR calibration assumes a Kroupa IMF, while the stellar masses were measured with a Chabrier IMF, so we apply a conversion from a Kroupa IMF to a Chabrier IMF (Kroupa = 1.122 Chabrier) to ensure consistency with the galaxy stellar mass measurements (e.g.~\citealt{Cimatti_2008}). 
While the \citet{Gilbank2010} relation is derived for lower-redshift objects, \citet{Sobral_2012} and \citet{Hayashi_2013} demonstrate that H$\alpha$ and [\ion{O}{II}] luminosities correlate well at higher-redshifts ($z \sim 1.5$), though indications are that galaxies are somewhat less dust extincted for a given H$\alpha$ luminosity compared with low redshift. As long as the abundance and properties of dust are not environment dependent, this should not alter our conclusions about the relative SFR in cluster and field galaxies. However, if the average dust content of cluster galaxies is lower than that of field galaxies (as is hinted by \citealt{McGee_2010,Zeimann_2013}), we would expect lower intrinsic SFRs for cluster galaxies (see also \citealt{Gallazzi_2009}). 

We discuss this further in Section~\ref{sec:Results}. We have checked that SFRs derived from [\ion{O}{II}] correlate with SFRs derived from H$\alpha$ for a very small subset of the [\ion{O}{II}] detections for which H$\alpha$ measurements are available from {\it HST}/WFC3 G141 grism spectroscopy from \citet{Matharu_2019}. We also note that the [\ion{O}{II}]-derived GOGREEN star-forming main sequence and the distributions of SFR with respect to the star-forming main sequence are remarkably similar to those from an H$\alpha$-derived galaxy sample at a similar epoch (we refer the reader to Section~\ref{sec:environmental_dependence_discussion} and Appendix~\ref{sec:appendix_Zeimann_comparison} for further details)

\begin{figure}
	\includegraphics[width=1.0\columnwidth]{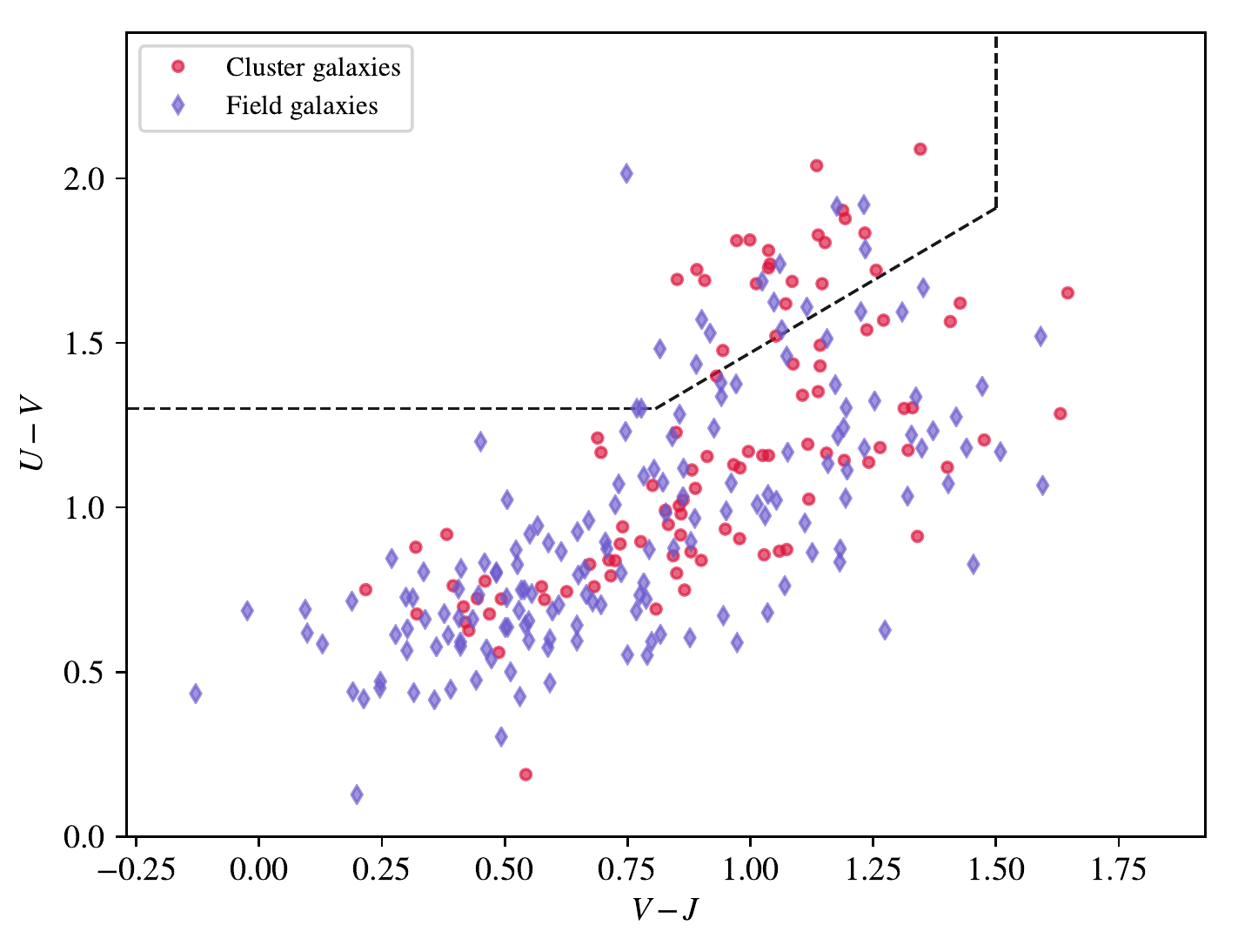}
    \caption{$UVJ$-diagram of [\ion{O}{II}] emitters with classifications from \citet{Muzzin_2013b} for $1 < z < 4$ (adapted from \citealt{Williams_2009}) where quiescent galaxies are above the dashed line in the upper left region, and star-forming galaxies are below the dashed line.}
    \label{fig:UVJ_80pc_flux_lim_flux_cal}
\end{figure}

In Figure~\ref{fig:UVJ_80pc_flux_lim_flux_cal} we show that this sample of star-forming galaxies has predominantly blue colours in $UVJ$ colour-colour space. We note that making an additional selection to exclude red galaxies from the sample does not qualitatively change our results. 

\subsection{Cluster and field sample properties}
\label{sec:cluster_field_properties}
\begin{figure}
	\includegraphics[width=1.0\columnwidth]{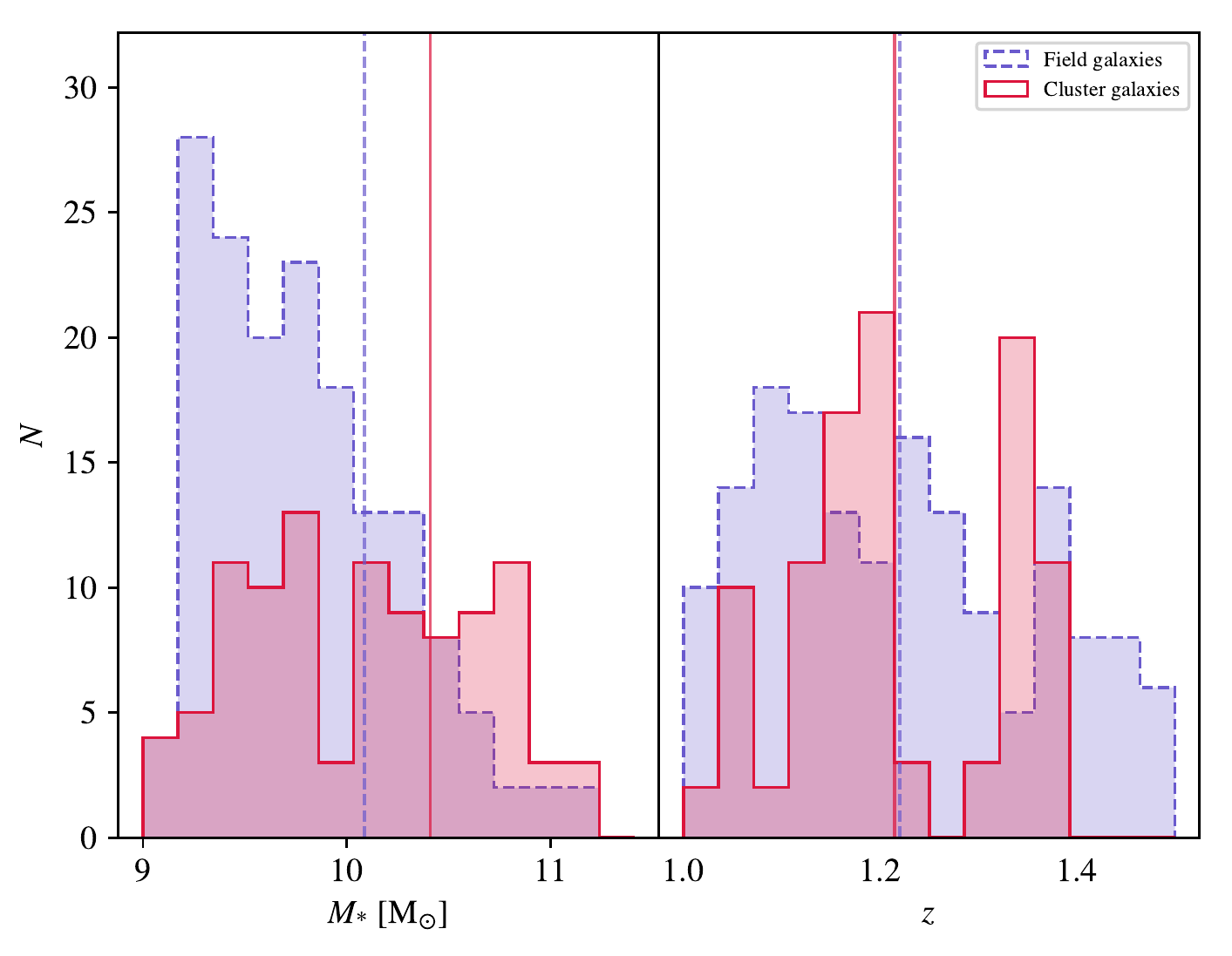}
    \caption{The stellar mass (\emph{left}) and redshift (\emph{right}) distributions for the field (dashed purple) and cluster (solid crimson) samples. The vertical lines represent the mean stellar mass and mean redshift of the cluster and field galaxy populations.}
    \label{fig:joint_z_hist_log_Mstel_80pc_flux_lim_flux_cal_narrow_bins}
\end{figure}

In Figure~\ref{fig:joint_z_hist_log_Mstel_80pc_flux_lim_flux_cal_narrow_bins} we present the stellar mass and redshift distributions of the cluster and field sample. In this [\ion{O}{II}]-emitting galaxy sample, field galaxies generally sit at lower stellar masses than the cluster galaxies. For this reason, we perform the subsequent analyses in bins of stellar mass. In Figure~\ref{fig:joint_z_hist_log_Mstel_80pc_flux_lim_flux_cal_narrow_bins}, it is also clear that while the mean redshifts of the two samples are very similar ($\Delta<z>=0.01$), the shapes of the distributions are quite different. 
The cluster galaxies are situated in the redshift space of individual clusters, whereas the field galaxies span a wider and more homogeneous redshift distribution \footnote{A two-sample KS test rejects that these two samples are drawn from the same stellar mass distribution with a $p$-value of 0.00017. The two-sample KS test $p$-value for the field and cluster redshift samples is found to be 0.06.}.

To ensure any difference in the star-forming main sequence between cluster and field is not due to differences in the underlying redshift distribution between our cluster and field samples, we apply a correction to the mean field SFR according to the mean redshift difference in cluster and field in each stellar mass bin. The correction is calculated using the observed cosmic star formation redshift relation for field galaxies from 
\citet{Schreiber_2015}:
\begin{multline}
  \log_{10}({\rm SFR}_{\rm MS}[{\rm M}_{\odot}~{\rm yr}^{-1}])
  = \\
  m-m_{0}+a_{0}r - a_{1}[{\rm max}(0,m-m_{1}-a_{2}r)]^{2},
  \label{eq:Schreiber_2015_SFR}
\end{multline}
where $r\equiv\rm{log_{10}}(\rm{1}+\it{z})$, $m\equiv\log_{10}(M_{*}/10^{9}~{\rm M}_{\odot})$, with $m_{\rm 0}=0.5\pm 0.07$, $a_{\rm 0}=1.5\pm 0.15$, $a_{\rm 1}=0.3\pm 0.08$, $m_{\rm 1}=0.36\pm 0.3$, and $a_{\rm 2}=2.5\pm 0.6$.
We note that the size of this correction is smaller than the statistical error bars in Figure~\ref{fig:log_Mstel_vs_SFR_binned_80pc_flux_lim_flux_cal_z_corr_schreiber15_bootstrap_lines}.
\section{Results}
\label{sec:Results}
\begin{figure}
	\includegraphics[width=1.0\columnwidth]{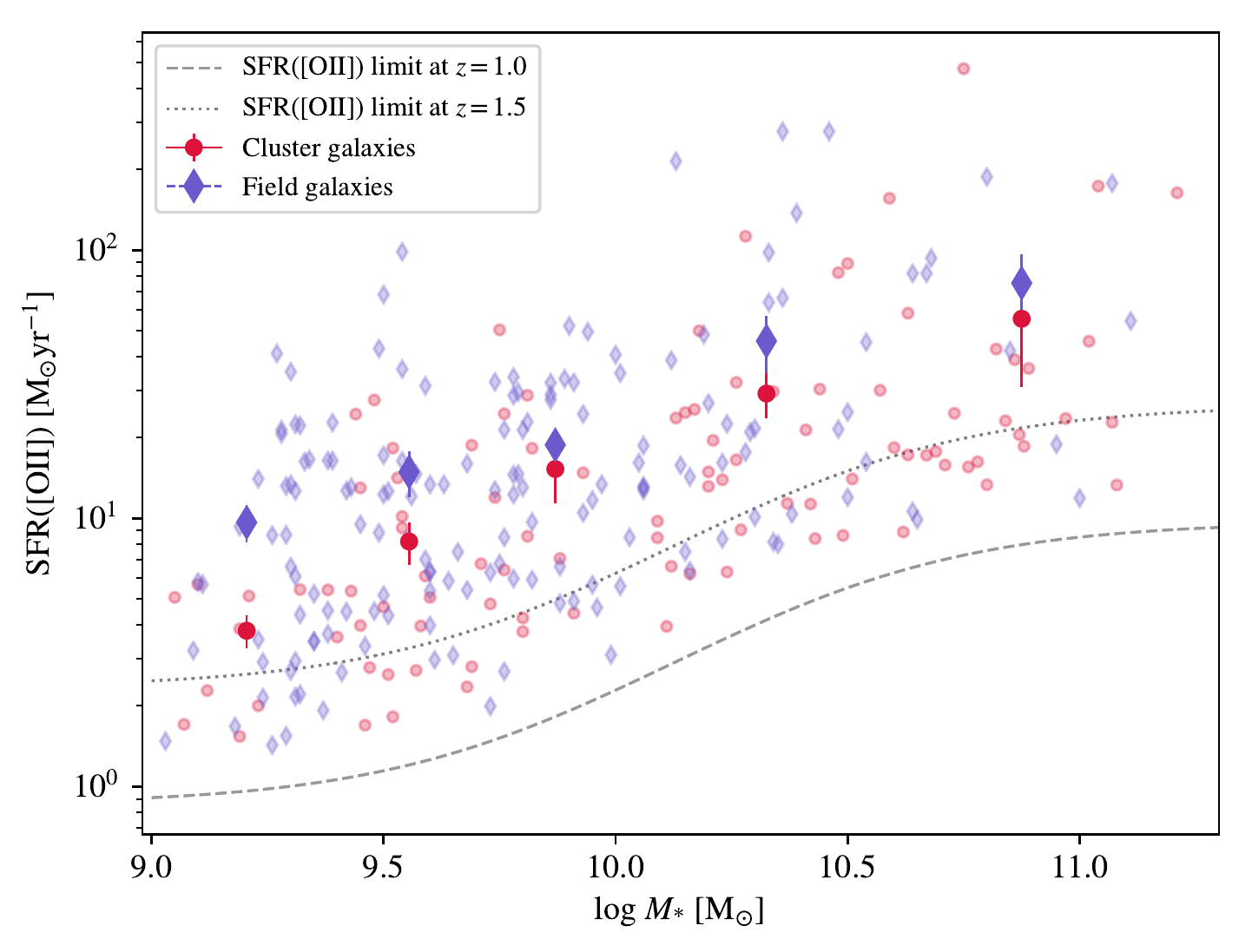}
    \caption{The main sequence of star formation of cluster galaxies versus field galaxies in the GOGREEN fields. The solid purple and crimson markers signify the mean field SFRs and the cluster galaxy SFRs in each stellar mass bin respectively. The field SFRs have been corrected using the cosmic SFR vs. $z$ relation of Equation~\ref{eq:Schreiber_2015_SFR} in order to match the mean redshift of cluster galaxies within each stellar mass bin. The error bars represent the bootstrap standard error from bootstrap resampling the data within each bin. The dashed and dotted grey lines represent the SFRs that correspond to the $80\%$ flux completeness limit at $z=1.0$ and $z=1.5$ respectively.}
    \label{fig:log_Mstel_vs_SFR_binned_80pc_flux_lim_flux_cal_z_corr_schreiber15_bootstrap_lines}
\end{figure}

In Figure~\ref{fig:log_Mstel_vs_SFR_binned_80pc_flux_lim_flux_cal_z_corr_schreiber15_bootstrap_lines}, we present the star-forming galaxy main sequence for cluster galaxies in the GOGREEN sample (crimson circles) and galaxies in the field (purple diamonds). We show the mean SFR of cluster and field galaxies (solid, larger datapoints with errorbars) in bins of stellar mass where bin widths are chosen adaptively to maintain a similar number of objects in each stellar mass bin. The error bars represent the bootstrap standard error.

From this comparison, we identify a modest environmental dependence on the star-forming galaxy main sequence: cluster galaxy SFRs are lower than their counterparts in the field at fixed stellar mass. To quantify this difference, we first fit the observed main sequence  for the full sample using the Theil-Sen estimator \citep{Theil_1950, Sen_1968} for robust linear regression of the relation  between $\log(M_{*})$ and $\log({\rm SFR})$. We then use this main sequence relation to calculate a $\Delta{\rm SFR}_{\rm MS}$ distribution for both the cluster galaxy and the field galaxy samples. 

\begin{figure}
	\includegraphics[width=1.0\columnwidth]{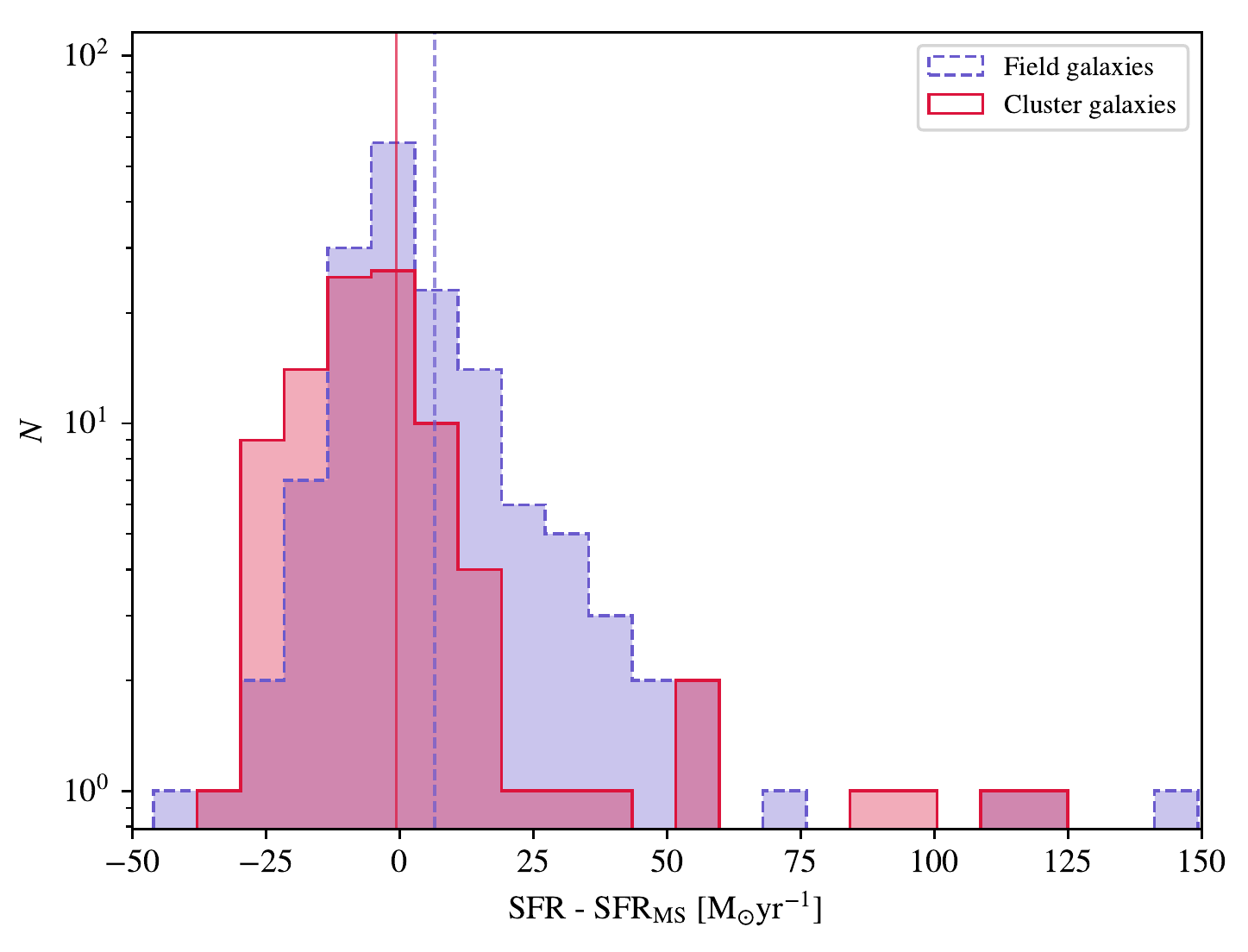}
    \caption{Cluster and field $\Delta{\rm SFR}_{\rm MS}$ distributions. The solid crimson vertical line represents the mean cluster galaxy $\Delta{\rm SFR}_{\rm MS}$ and the dashed purple vertical line represents the mean field galaxy $\Delta{\rm SFR}_{\rm MS}$.}
    \label{fig:delta_MS_SFR_hist_80pc_flux_lim_flux_cal_SFR_lim_1.0_z_1.5_SFR_zlim_1.0}
\end{figure}

The cluster and field $\Delta{\rm SFR}_{\rm MS}$ distributions are shown in Figure~\ref{fig:delta_MS_SFR_hist_80pc_flux_lim_flux_cal_SFR_lim_1.0_z_1.5_SFR_zlim_1.0}, where the solid crimson vertical line represents the mean cluster galaxy $\Delta{\rm SFR}_{\rm MS}$ and the dashed purple vertical line represents the mean field galaxy $\Delta{\rm SFR}_{\rm MS}$. The mean difference in $\log(\Delta{\rm SFR}_{\rm MS})$ between the cluster and field\footnote{We exclude one cluster galaxy and three field galaxies whose $\Delta{\rm SFR}_{\rm MS}$ is more than 2$\sigma$ outside this fit to avoid these extreme values of $\Delta{\rm SFR}$ from dominating the comparison of different populations.} is $-0.145~{\rm M}_{\odot}~{\rm yr}^{-1}$. 

Dividing by the combined bootstrap standard error (0.045 dex), yields a significance of $\sim3.3\sigma$ \footnote{If we downsample the field to match that of the cluster galaxy sample size, we find similar values of significance (for 10,000 random subsamples, the median and mean significance is $2.8\sigma$).}. A two-sample KS test rejects that these two samples come from the same distribution with a $p$-value of $1.4 \times 10^{-5}$. We see a similar difference in the specific SFRs of the samples, with a difference in average $\log({\rm sSFR})$ of $-0.128\pm 0.046$ dex at the $\sim2.8\sigma$ level. We now focus on the shape of the $\Delta{\rm SFR}_{\rm MS}$ distributions.  We see that the cluster population has a small tail to lower SFRs. The most significant difference is the near absence of cluster galaxies with significantly enhanced SFRs, although these galaxies are common in the field.  

We note that the small correction we make to account for the different mean redshifts of the two samples using the \citet{Schreiber_2015} as described in Section~\ref{sec:discussion} does not have a significant effect on these results.  However, the fixed [\ion{O}{II}] flux limit corresponds to a different SFR limit at $z=1.0$ and $z=1.5$.  Because of the different redshift distributions of the cluster and field samples, this can lead to a difference that is not accounted for by this correction. To be even more conservative, if we select a subsample for which the SFR is greater than that corresponding to the $80\%$ flux completeness limit at $z=1.5$, we reduce the sample to 64 cluster and 130 field galaxies, but find the same qualitative trend between cluster and field at the $\sim2.4\sigma$ level.
Separating the sample by redshift, we find that our result is driven by the lower redshift end of the sample. At $z < 1.3$, the significance of the difference between cluster and field $\log(\Delta{\rm SFR}_{\rm MS})$ and $\log(\Delta{\rm sSFR}_{\rm MS})$ is $\sim4.9\sigma$ and $\sim4.6\sigma$, respectively. Our sample above redshift $z > 1.3$ is small, limited to 21 cluster and 42 field galaxies. We find no significant difference between the cluster and field $\log(\Delta{\rm SFR}_{\rm MS})$ and $\log(\Delta{\rm sSFR}_{\rm MS})$ for this small subsample.\footnote{In these redshift subset comparisons, we apply a SFR limit derived by converting the $80\%$ $F([\ion{O}{II}])$ limit to a SFR at the higher-redshift limit of the sample -- i.e.~$z=1.3$ for the  $1.0<z<1.3$ sample and $z=1.5$ for the $1.3<z<1.5$ sample.}

\section{Discussion}
\label{sec:discussion}
\subsection{Environmental dependence of the star-forming main sequence}
\label{sec:environmental_dependence_discussion}
In this study, we identify an environmental dependence on the star-forming galaxy main sequence at $1.0 < z < 1.5$, where cluster galaxies have lower log $\Delta {\rm SFR}_{\rm MS}$ at fixed stellar mass than their counterparts in the field by a factor of 1.4, with a significance in log $\Delta {\rm SFR}_{\rm MS}$ of $\sim 3.3\sigma$ across all stellar masses, but strongest at lower stellar masses (Figure~\ref{fig:log_Mstel_vs_SFR_binned_80pc_flux_lim_flux_cal_z_corr_schreiber15_bootstrap_lines}). 

Our findings are in good agreement with those of \citet{Noirot_2018}, who find a significant suppression in the main sequence of the CARLA cluster sample, relative to the field sample of \citet{Whitaker_2014} at $z\sim1.5$. However, our results appear to differ somewhat from some other studies at a comparable redshift.  For example, \citet{Zeimann_2013}, use {\it HST} grism data to measure H$\alpha$ fluxes of galaxies in  18 clusters at $1.0<z<1.5$ and find no significant environmental dependence of the star-forming main sequence. 

Comparing their H$\alpha$-derived SF properties directly in a consistent manner with that of this paper, we find that the GOGREEN [\ion{O}{II}]-derived $\Delta{\rm SFR}_{\rm MS}$ distributions are very similar to that of the H$\alpha$-derived $\Delta{\rm SFR}_{\rm MS}$ from \citet{Zeimann_2013}. From a direct comparison of cluster and field $\Delta{\rm SFR}_{\rm MS}$ distributions, we see a hint that the cluster $\Delta{\rm SFR}_{\rm MS}$ are lower than the field at lower stellar masses, but in agreement with \citet{Zeimann_2013}, this difference is not statistically significant. We refer the reader to Appendix~\ref{sec:appendix_Zeimann_comparison} for more details. 
Although the authors do observe that the H$\alpha$ equivalent widths (EWs) of the cluster star-forming galaxies are lower than their counterparts in the field for  $M_{*}<10^{10}~{\rm M}_{\odot}$, they attribute this to a difference in SFH.  They also find weak evidence that the dust content of cluster galaxies is lower than that in the field. 

Using a sample of galaxy groups at $0.5 < z < 1.1$ from COSMOS, AEGIS, ECDFS, and CDFN fields, \citet{Erfanianfar_2016} find little variation in the star-forming MS with environment. Our result also differs from other studies at a range of redshifts \citep[e.g.][]{Elbaz_2007, Cooper_2008, Popesso_2011} which claim to observe a {\it reversal} of the sSFR-density trend, such that star-forming galaxies in dense environments have {\it higher} sSFR than the field. However, it is noted in these studies that the samples are dominated by groups rather than clusters, and so this reversal in the trend of sSFR-density does not necessarily apply to massive clusters.  Many of these results are also driven by higher mass galaxies, where dust corrections are most important.  We note that \citet{Popesso_2011} find a trend that is similar to the one we observe, when restricted to lower stellar mass galaxies. 

Taken by itself, the environmental dependence on the star-forming galaxy main-sequence that we find allows for numerous interpretations. In this work, we consider two possible interpretations: a formation time dependence on cluster and field populations, motivated by the results of van der Burg et al. in preparation and Webb et al. in preparation; and a delayed-than-rapid quenching model based on \citet{Wetzel_2013} which has shown to provide a good match to certain studies at lower redshift. While these scenarios are not necessarily the only interpretations possible given our observational results, they serve as useful reference points.

\subsection{Formation time predictions}
 \begin{figure}
	\includegraphics[width=1.0\columnwidth]{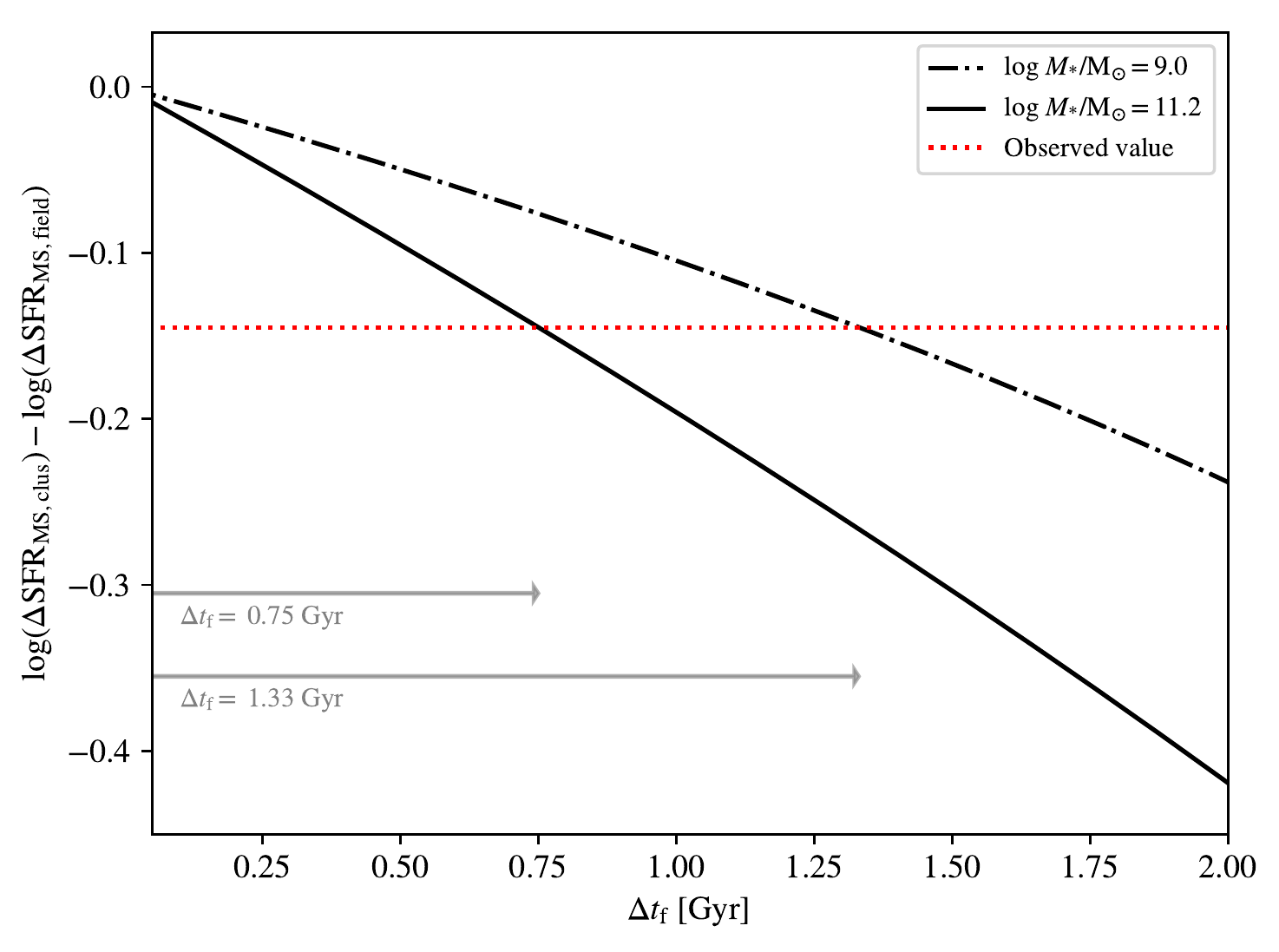}
    \caption{The average difference in formation time between cluster and field galaxies versus the difference in $\log(\Delta{\rm SFR}_{\rm MS})$ between the two samples. These values are derived from the \citet{Schreiber_2015} cosmic SFR vs. redshift relation assuming two fiducial stellar masses located at the minimum (dashed dotted) and maximum (solid) ends of the stellar mass distribution of our sample. The red dotted line shows the observed $\log(\Delta{\rm SFR}_{\rm MS})$ difference between cluster and field.}
    \label{fig:Screiber_2015_delta_t_vs_delta_log_SFR_z_final}
\end{figure}
A simple scenario to explain the observed suppression in the cluster star-forming galaxy main sequence compared to that of the field without the need to invoke environmentally-driven quenching in cluster environments is that cluster galaxies have simply formed earlier than field galaxies\footnote{This scenario is not necessarily supported by assembly bias, where the relation between  age and clustering depends on the halo mass relative to the characteristic collapse mass, $M_{\rm c}$, at a given redshift. At $M_{\rm vir} >> M_{\rm c}$, younger haloes are clustered more strongly than older haloes, with the reverse being true at $M_{\rm vir} << M_{\rm c}$ \citep{Wechsler_2006, Gao_2007, Zentner_2014}.}.
 
To explore this scenario, we employ the \citet{Schreiber_2015} cosmic SFR vs. redshift relation, to find the difference in redshift (and hence time) required to produce the observed difference between cluster and field $\log(\Delta{\rm SFR}_{\rm MS})$ of $-0.145~{\rm M}_{\odot}~{\rm yr}^{-1}$. In Figure~\ref{fig:Screiber_2015_delta_t_vs_delta_log_SFR_z_final} we show an example of the formation time difference between cluster and field galaxies versus the resulting difference in $\log(\Delta{\rm SFR}_{\rm MS})$ for two fiducial stellar masses which are taken as the minimum (dashed dotted) and maximum (solid) stellar masses of galaxies in our sample . The red dotted line shows the observed $\log(\Delta{\rm SFR}_{\rm MS})$ difference between cluster and field, which corresponds to formation time differences of $>0.75$ Gyr for high mass galaxies ($\log(M_{*} /M_{\odot}) \sim 11.2$) and $>1.3$ Gyr for low mass galaxies ($\log(M_{*} /M_{\odot}) \sim 9.0$). 

While these formation time differences are long, requiring a substantial ‘head start’ for cluster galaxies compared to galaxies in the field, they are not inconsistent with works focussing on other observables such as the fundamental plane, for example, \citet{Saglia_2010}, who find differences in ages of cluster and field galaxies of $\sim$1 Gyr at a fixed stellar mass and redshift using the EDisCS cluster sample. However, we note that other works also based on fundamental plane such as \citet{van_Dokkum_2007} find smaller differences in the ages of stars in massive cluster galaxies compared to the field of $\sim$0.4 Gyr.

\subsection{Environmental quenching timescale predictions}
\label{sec:predicted_quenching_timescales}
We now move to exploring an alternative scenario where a simple interpretation of our observations is that recently accreted cluster galaxies are undergoing an environmentally-driven decline in star formation without the need to invoke a formation time difference between cluster and field galaxies. To try and quantify what the implied quenching rates would be, we consider a toy model based on \cite{Wetzel_2013}. In this model, after a satellite galaxy infalls into a cluster halo, there is a period of time referred to as the `delay-time', $t_{\rm delay}$, within which a galaxy's SFR follows that of the typical field evolution. After the delay-time, there is then a period of rapid decrease in SFR which declines at a rate defined by $\tau$, often referred to as the `fading time':
\begin{equation}
  {\rm SFR}=\begin{cases}
     {\rm SFR}(t_{\rm start})e^{(-(t - t_{\rm start})/\tau)}, & \text{$t> t_{\rm start}$}\\
     {\rm SFR}(t), & \text{$t\leq t_{\rm start}$} \; ,
  \end{cases}
  \label{eq:SFR_quenching_modelling}
\end{equation}
\noindent where $t_{\rm start} =  t_{\rm infall} + t_{\rm delay}$.

This `delayed-then-rapid' quenching scenario is also supported by studies such as \citet{McCarthy_2008} who find that satellite galaxies in hydro-dynamical simulations typically maintain a significant fraction of their hot gas after infall into the cluster potential. \citet{Mok_2014}, who study the Group Environment Evolution Collaboration 2 (GEEC2) sample of galaxy groups at $0.8 < z <1.0$, also find that it is necessary to invoke a model that includes a period of typical field SF activity before rapidly quenching to explain the observed fractions of star-forming/intermediate/quiescent fractions
(a no delay scenario would require longer fading times which would overproduce intermediate-colour galaxies).

It is our goal to constrain the parameters $t_{\rm delay}$ and $\tau$ in this model using the measured properties of cluster and field galaxies. We choose to focus on galaxies in both the observed cluster and field population with log $M_{*}/\rm{M_{\odot}}<10.3$ and at $1.0 < z < 1.3$ in an effort to restrict our study to satellite galaxies where we expect quenching is not dominated by internal processes (unlike massive galaxies), and where our [\ion{O}{II}]-derived SFRs are expected to be most robust given the lower dust extinction.

\begin{figure}
	\includegraphics[width=1.0\columnwidth]{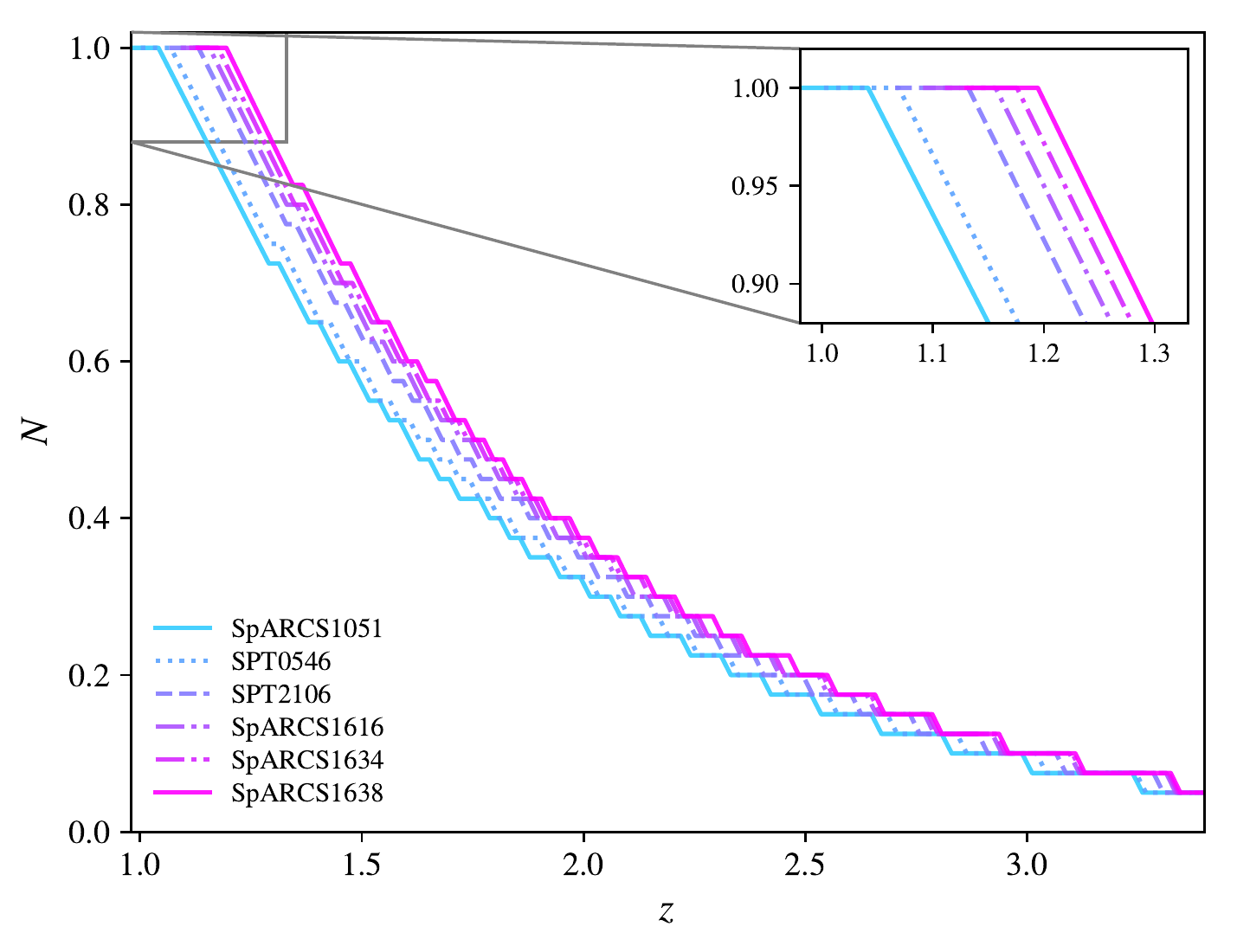}
    \caption{Cumulative distributions of physically-motivated accretion redshifts for the modelled galaxies based on six clusters at $z< 1.3$, described in further in detail in Section~\ref{sec:predicted_quenching_timescales}. The location where the curves plateau corresponds to the observed cluster redshift.}
    \label{fig:cluster_accretion_z_cumulative_dist}
\end{figure}

We first generate a parent sample of galaxies whose infall redshifts correspond to the time at which the galaxies were accreted into the cluster. We employ a physically-motivated distribution of infall redshifts, following \citet{Neistein_2006, Neistein_2008} and use a functional form of the average mass accretion history similar to that of the main progenitor (MP) halo in the Extended Press-Schechter formalism \citep{Press_1974, Bond_1991, Lacey_1993}\footnote{In this model, mass accretion history is based on when a galaxy is accreted into the most massive halo i.e, this model does not consider the effect of group preprocessing. Alternatively, if massive accretion is based on when galaxy first becomes a satellite, we would expect longer delay timescales.}. For more details regarding how this infall redshift distribution was generated, we refer the reader to Appendix~\ref{sec:appendix_cluster_infall}. We show the cumulative distribution of accretion redshifts for the modelled galaxies for each of the six clusters at $z<1.3$ in Figure~\ref{fig:cluster_accretion_z_cumulative_dist}. 

We use the observed field sample, with 50 random selections per cluster member and evolve their SFRs to these infall redshifts, via the \citet{Schreiber_2015} relation described in Section~\ref{sec:cluster_field_properties}. To be able to compare  this model cluster population to the observed cluster distribution, we then evolve their SFRs from the assigned infall redshifts forward to the parent cluster redshift according to a range of $t_{\rm delay}$ and $\tau$ timescales. Our aim is to exclude unlikely combinations of $t_{\rm delay}$ and $\tau$ by deducing which resulting SFR distributions are significantly different from the observed cluster galaxy SFR distribution\footnote{Note that by construction, if there is no environmental quenching, the model cluster SFR distribution should end up looking exactly like the observed field.}. 

We use the main sequence fit described in Section~\ref{sec:Results} to calculate a $\Delta SFR_{\rm clus}$ distribution for both the observed cluster galaxies and the model cluster population, $\Delta SFR_{\rm model\;clus}$, for all timescales. For each timescale realisation, we then perform a two-sample KS test on the observed cluster galaxy distribution and the model cluster population distribution i.e., KS($\Delta SFR_{\rm clus}$, $\Delta SFR_{\rm model\;clus}$) to test against the null hypothesis that two independent samples are drawn from the same continuous distribution. Higher two-sample KS test $p$-values indicate a smaller absolute maximum distance between the cumulative distribution functions of the two samples, and therefore more likely timescale values.

In addition to the parameterisation of the delayed-then-rapid quenching model described in Equation~\ref{eq:SFR_quenching_modelling}, we also present an alternative parameterisation following from \citet{Hahn_2017}\footnote{In \citet{Hahn_2017}, this parameterisation is adopted for central galaxies, while here we use this parameterisation for satellite galaxies.} where:
\begin{equation}
  {\rm SFR}=\begin{cases}
     {\rm SFR}(t)e^{(-(t - t_{\rm start})/\tau_{\rm H})}, & \text{$t> t_{\rm start}$}\\
     {\rm SFR}(t), & \text{$t\leq t_{\rm start}$} \; .
  \end{cases}
  \label{eq:Hahn_SFR_quenching_modelling}
\end{equation}

In the first parameterisation, Equation~\ref{eq:SFR_quenching_modelling}, $\tau$ represents the SFR evolution during the quenching epoch ($t > (t_{\rm infall} + t_{\rm delay}$)). In the absence of environmental effects, $\tau < \infty$ as the SFRs of the field galaxies evolve\footnote{Analytically, in the absence of environmental effects, $\tau$ exists only if $t> t_{\rm start}$, but when fitting the model to data, there is a degeneracy between a long $t_{\rm delay}$ and a $\tau$ equal to the effective SFR decline of the field population.}. In the second parameterisation, Equation~\ref{eq:Hahn_SFR_quenching_modelling}, $\tau_{\rm H}$ represents the additional quenching on top of the normal SFR evolution, and is therefore a clearer indicator of the role of environmental quenching. We opt to present results of both models to allow us to compare with different environmental quenching timescales presented in the literature.

\begin{figure}
	\includegraphics[width=1.0\columnwidth]{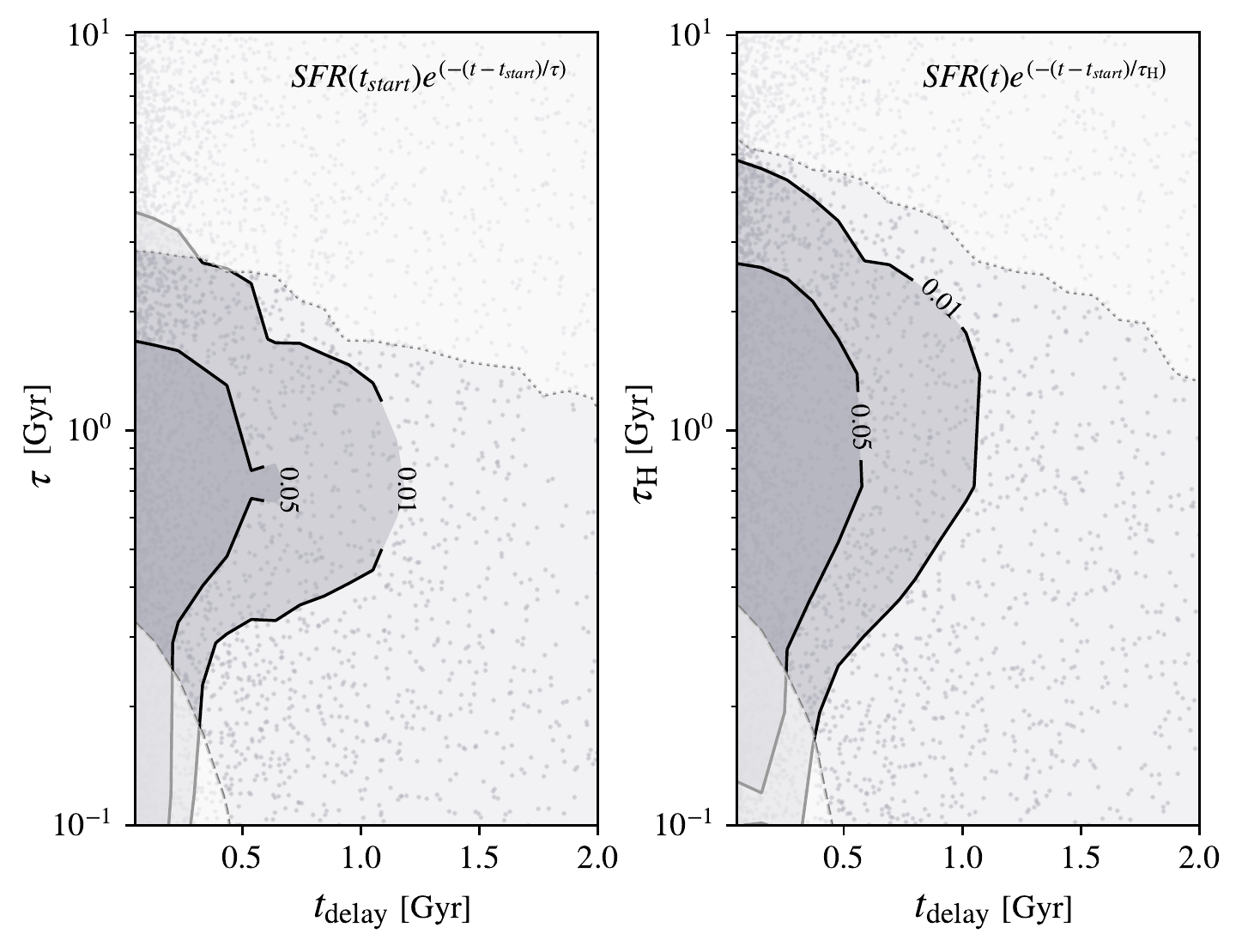}
    \vspace{-4mm}\caption{In this figure we present density contours showing the likely range of quenching timescale parameters: $t_{\rm delay}$ and $\tau$, adopted in the toy model derived from the observed difference between the GOGREEN cluster galaxy and field star-forming galaxy main sequence. The density contours are derived from the two-sample KS-test $p$-values comparing the observed GOGREEN cluster galaxy SFR distribution and the predicted field distribution. Note that these distributions have been subtracted from the main sequence relation for cluster and field galaxies. Lower two-sample KS test $p$-values indicate more absolute maximum distance between the cumulative distribution functions of the two samples, and therefore more less timescale values. The left subplot shows the quenching timescales based on Model 1 from Equation~\ref{eq:SFR_quenching_modelling}, and the right subplot shows quenching timescales based off of Model 2 from Equation~\ref{eq:Hahn_SFR_quenching_modelling}. The opaque region in the lower corner below the dashed gray line signifies the timescale parameter space where the fraction of model star-forming galaxies that drop below the $F([\ion{O}{II}])$ limit ($F([\ion{O}{II}])=2.2\times10^{-17}~{\rm erg}~{\rm cm}^{-2}~{\rm s}^{-1}$), $f_{\rm dropout}$, is $> 0.8$, and the  opaque region in the upper corner above the dotted gray line signifies the timescale parameter space where $f_{\rm dropout}$, is $ <0.1$.}
    \label{fig:t_delay_vs_tau_subplots_KS_pval_paper}
\end{figure}

In Figure~\ref{fig:t_delay_vs_tau_subplots_KS_pval_paper}, we show the resulting quenching timescale parameter space density contours for 10,000 simulations of different $t_{\rm delay}$ and $\tau$ produced with the two parameterisations. Examples of resulting $<\Delta \rm{log}\; SFR_{\rm MS}>$ distributions for specific $t_{\rm delay}$ and $\tau$ timescales can be found in Figure~\ref{fig:model_delta_MS_SFR_hist_specific_examples_i_4_1.0_z_1.3_SFR_zlim_1.0} in the Appendix. The opaque region in the lower corner below the dashed gray line signifies the timescale parameter space where the fraction of model star-forming galaxies that drop below the $F([\ion{O}{II}])$ limit ($F([\ion{O}{II}])=2.2\times10^{-17}~{\rm erg}~{\rm cm}^{-2}~{\rm s}^{-1}$), $f_{\rm dropout}$, is $> 0.8$, and the opaque region in the upper corner above the dotted gray line signifies the timescale parameter space where  $f_{\rm dropout}$, is $ <0.1$. Given that quenched fraction excesses are unlikely to be as high as 0.8 or as low as 0.1 (van der Burg et al. in preparation), we see that a combination of a short delay and short fade time is unlikely given this conservative dropout fraction. If, in reality, the quenched fraction is lower than that assumed here, we would expect a further restriction in this region in quenching timescale parameter space.
    
We constrain $t_{\rm delay} < 1.2$ Gyr at the 99$\%$ level. For very rapid quenching scenarios, the constraint on the delay time is stronger ($t_{\rm delay} < 0.25$ Gyr at the 99$\%$ level). We find that the timescale for environmental quenching is $\tau_{\rm H} < 6$ Gyr, allowing for modest environmental quenching, as long as delay times are reasonably short, so that a significant population of galaxies are affected. 
At face value, given the broad constraints shown in Figure~\ref{fig:t_delay_vs_tau_subplots_KS_pval_paper} on $t_{\rm delay}$, we cannot rule out a slow-quenching scenario where galaxies quench slowly from infall at a rate that is accelerated somewhat in comparison to the field.  The timescales we constrain are in general agreement with works such as \citet{Muzzin_2014} and \citet{Mok_2014}, who derive constraints using galaxy cluster phase-space and fractions of red, blue and green galaxies respectively in samples of clusters at $0.8 < z< 1.0 $. However, there is a hint that our short delay timescale predictions could be in tension with some works that favour longer delay times. For example, \citet{Balogh_2016} find delay times of $\sim 4.5$ Gyr at $z=0$ for galaxies in a similar mass range (assuming fixed fading time of $\sim 0.5$ Gyr), which would correspond to $\sim1.3$ Gyr if the delay time scales with the dynamical time.

We note again that different timescale combinations would lead to very different quenched fractions. For example, short delay times combined with short fade times would result in almost all cluster galaxies being quenched, while very long delay times combined with long tau would lead to implausibly small quenched fractions.  Combining the constraints from this work with our analysis of the stellar mass functions (van der Burg et al. in preparation; Reeves et al. in preparation) will lead to much tighter constraints on these timescales.

An important caveat of this delay timescale modelling is that we assume that dust content of cluster and field galaxies are the same, and that the difference we observe between cluster and field galaxies is due to a difference in SFH. \citet{Zeimann_2013} find that at fixed stellar mass, field star-forming galaxies have slightly higher extinction on average than star-forming galaxies in clusters, (though the difference is within their error bars). If dust extinction is higher for field galaxies compared to cluster galaxies in this study, the difference in the star-forming galaxy main sequence would be underestimated, and we would expect shorter delay times and/or longer fading times than those predicted by this toy model.

We also note it is likely that our cluster sample contains a certain number of interloping galaxies that are not gravitationally bound to the cluster but are mistaken for cluster members. Even for large, well sampled cluster galaxy spectroscopic samples, the interloper rate for dynamical membership techniques is predicted to be at least 15$\%$ \citep{Duarte_2015, Wojtak_2018}. The impact of interloping galaxies is expected to dilute any environmental dependence of the star-forming galaxy main sequence, which would also result in a larger difference in the MS between cluster and field and therefore require shorter delay times and/or longer fading times than those predicted by this toy model.

\section{Conclusions}
\label{sec:conclusions}
In this paper we explore the environmental dependence of the star-forming galaxy main sequence in an unprecedented sample of homogeneously selected deep spectroscopic observations of galaxies in 11 galaxy cluster fields at $1.0 < z < 1.5$ from the GOGREEN survey. Our major findings can be summarised as follows:
\begin{enumerate}
    \item We take a Bayesian model approach in detecting [\ion{O}{II}] emission from the GOGREEN galaxy spectra, employing BIC to distinguish cases where there is strong evidence supporting a model with [\ion{O}{II}] emission over a model with no emission, taking into account the noise properties of the continuum in each individual spectrum. Employing a conservative F([\ion{O}{II}]) limit, we detect [\ion{O}{II}] emission in 100 cluster galaxies and 162 field galaxies across 11 of the GOGREEN cluster fields.
     \item When accounting for differences between cluster and field redshift properties, we find that the cluster galaxy main sequence is lower compared to that of the field galaxy main sequence at $1.0 < z < 1.5$, with a difference of $\sim3.3\sigma$. We find that this result is driven by the lower redshift end of the sample, and is more significant for lower stellar mass galaxies.
     \item This observed environmental dependence on the star-forming galaxy main sequence allows for numerous interpretations. We explore several of these scenarios, placing constraints based on our measurements. One such interpretation is that cluster galaxies are simply formed earlier than field galaxies. Given our observations, long formation time differences between the two populations of $> 0.8$ Gyr would be required in this interpretation. Focussing on an alternative scenario whereby environmentally-induced quenching occurs, we model the likely quenching timescales given the size of the observed difference between the cluster and field main sequence, and find that our data favour delay times of $< 1.2$ Gyr at the 99$\%$ level.
\end{enumerate}   

\noindent The formation time and star formation quenching timescale constraints from this work will be combined with analysis of the stellar mass functions of galaxy clusters (van der Burg et al. in preparation), galaxy groups (Reeves et al. in preparation) and quiescent galaxy stellar population ages (Webb et al. in preparation) from the GOGREEN survey, providing tighter constraints on models of galaxy evolution in dense environments.

\section*{Acknowledgements}
We thank Emiliano Munari for providing us with the C.L.U.M.P.S. algorithm in advance of publication. We also thank Jasleen Matharu and Gregory Zeimann for providing H$\alpha$ measurements.  This research is supported by the following grants:  European Space Agency (ESA) Research Fellowship (LJO); NSERC
Discovery grants (MLB and HKCY); Canada Research Chair program and the Faculty of Arts and Science, University of Toronto (HKCY); NSF
grants AST-1517815 and AST-1211358 (GHR), AST-1517863
(GW) and AST-1518257 (MCC); NASA, through grants AR14310.001 and GO-12945.001-A (GHR), GO-13306, GO-13677, GO-13747 and GO-13845/14327 (GW), AR-13242 and AR-14289
(MCC); STFC (SLM); the Chilean Centro de Excelencia en Astrof\'isica y Tecnolog\'ias Afines (CATA) BASAL grant AFB-170002 (RD); Universidad Andr\'es Bello Internal Project DI-12-19/R (JN); the National Research Foundation of South Africa (DGG);
and ALMA-CONICYT grant 31180051 (PC).

This paper includes data gathered with the Gemini Observatory,
which is operated by the Association of Universities for Research
in Astronomy, Inc., under a cooperative agreement with the NSF
on behalf of the Gemini partnership: the National Science Foundation (United States), the National Research Council (Canada),
CONICYT (Chile), Ministerio de Ciencia, Tecnologa e Innovacin
Productiva (Argentina), and Ministrio da Ciłncia, Tecnologia e
Inovao (Brazil); the 6.5 metre Magellan Telescopes located at Las
Campanas Observatory, Chile; the Canada–France–Hawaii Telescope (CFHT) which is operated by the National Research Council of Canada, the Institut National des Sciences de l’Univers of
the Centre National de la Recherche Scientifique of France, and
the University of Hawaii; MegaPrime/MegaCam, a joint project of
CFHT and CEA/DAPNIA; Subaru Telescope, which is operated by the National Astronomical Observatory of Japan; and the ESO
Telescopes at the La Silla Paranal Observatory under programme
ID 097.A-0734. This research made use of Astropy \citep{astropy_2013, astropy_2018}, SciPy \citep{SciPy_2019}, {\sc emcee} \citep{Foreman-Mackey_2013}, Matplotlib \citep{Hunter_2007} and NumPy \citep{Oliphant_2015}.




\bibliographystyle{mnras}
\bibliography{GOGREEN_OII_SFR.bib} 

\begin{thebibliography}{}
\makeatletter
\relax
\def\mn@urlcharsother{\let\do\@makeother \do\$\do\&\do\#\do\^\do\_\do\%\do\~}
\def\mn@doi{\begingroup\mn@urlcharsother \@ifnextchar [ {\mn@doi@}
  {\mn@doi@[]}}
\def\mn@doi@[#1]#2{\def\@tempa{#1}\ifx\@tempa\@empty \href
  {http://dx.doi.org/#2} {doi:#2}\else \href {http://dx.doi.org/#2} {#1}\fi
  \endgroup}
\def\mn@eprint#1#2{\mn@eprint@#1:#2::\@nil}
\def\mn@eprint@arXiv#1{\href {http://arxiv.org/abs/#1} {{\tt arXiv:#1}}}
\def\mn@eprint@dblp#1{\href {http://dblp.uni-trier.de/rec/bibtex/#1.xml}
  {dblp:#1}}
\def\mn@eprint@#1:#2:#3:#4\@nil{\def\@tempa {#1}\def\@tempb {#2}\def\@tempc
  {#3}\ifx \@tempc \@empty \let \@tempc \@tempb \let \@tempb \@tempa \fi \ifx
  \@tempb \@empty \def\@tempb {arXiv}\fi \@ifundefined
  {mn@eprint@\@tempb}{\@tempb:\@tempc}{\expandafter \expandafter \csname
  mn@eprint@\@tempb\endcsname \expandafter{\@tempc}}}

\bibitem[\protect\citeauthoryear{{Abazajian} et~al.,}{{Abazajian}
  et~al.}{2009}]{Abazajian_2009}
{Abazajian} K.~N.,  et~al., 2009, \mn@doi [\apjs]
  {10.1088/0067-0049/182/2/543}, \href
  {https://ui.adsabs.harvard.edu/abs/2009ApJS..182..543A} {182, 543}

\bibitem[\protect\citeauthoryear{{Ashman}, {Bird}  \& {Zepf}}{{Ashman}
  et~al.}{1994}]{Ashman_1994}
{Ashman} K.~M.,  {Bird} C.~M.,   {Zepf} S.~E.,  1994, \mn@doi [\aj]
  {10.1086/117248}, \href
  {https://ui.adsabs.harvard.edu/abs/1994AJ....108.2348A} {108, 2348}

\bibitem[\protect\citeauthoryear{{Astropy Collaboration} et~al.,}{{Astropy
  Collaboration} et~al.}{2013}]{astropy_2013}
{Astropy Collaboration} et~al., 2013, \mn@doi [\aap]
  {10.1051/0004-6361/201322068}, \href
  {https://ui.adsabs.harvard.edu/abs/2013A%26A...558A..33A} {558, A33}

\bibitem[\protect\citeauthoryear{{Astropy Collaboration} et~al.,}{{Astropy
  Collaboration} et~al.}{2018}]{astropy_2018}
{Astropy Collaboration} et~al., 2018, \mn@doi [\aj] {10.3847/1538-3881/aabc4f},
  \href {https://ui.adsabs.harvard.edu/abs/2018AJ....156..123A} {156, 123}

\bibitem[\protect\citeauthoryear{{Baldry}, {Glazebrook}, {Brinkmann},
  {Ivezi{\'c}}, {Lupton}, {Nichol}  \& {Szalay}}{{Baldry}
  et~al.}{2004}]{Baldry_2004}
{Baldry} I.~K.,  {Glazebrook} K.,  {Brinkmann} J.,  {Ivezi{\'c}} {\v{Z}}.,
  {Lupton} R.~H.,  {Nichol} R.~C.,   {Szalay} A.~S.,  2004, \mn@doi [\apj]
  {10.1086/380092}, \href
  {https://ui.adsabs.harvard.edu/abs/2004ApJ...600..681B} {600, 681}

\bibitem[\protect\citeauthoryear{{Balogh} et~al.,}{{Balogh}
  et~al.}{2004a}]{Balogh_2004a}
{Balogh} M.,  et~al., 2004a, \mn@doi [\mnras]
  {10.1111/j.1365-2966.2004.07453.x}, \href
  {https://ui.adsabs.harvard.edu/abs/2004MNRAS.348.1355B} {348, 1355}

\bibitem[\protect\citeauthoryear{{Balogh}, {Baldry}, {Nichol}, {Miller},
  {Bower}  \& {Glazebrook}}{{Balogh} et~al.}{2004b}]{Balogh_2004b}
{Balogh} M.~L.,  {Baldry} I.~K.,  {Nichol} R.,  {Miller} C.,  {Bower} R.,
  {Glazebrook} K.,  2004b, \mn@doi [\apjl] {10.1086/426079}, \href
  {https://ui.adsabs.harvard.edu/abs/2004ApJ...615L.101B} {615, L101}

\bibitem[\protect\citeauthoryear{{Balogh} et~al.,}{{Balogh}
  et~al.}{2016}]{Balogh_2016}
{Balogh} M.~L.,  et~al., 2016, \mn@doi [\mnras] {10.1093/mnras/stv2949}, \href
  {https://ui.adsabs.harvard.edu/abs/2016MNRAS.456.4364B} {456, 4364}

\bibitem[\protect\citeauthoryear{{Balogh} et~al.,}{{Balogh}
  et~al.}{2017}]{Balogh_2017}
{Balogh} M.~L.,  et~al., 2017, \mn@doi [\mnras] {10.1093/mnras/stx1370}, \href
  {https://ui.adsabs.harvard.edu/abs/2017MNRAS.470.4168B} {470, 4168}

\bibitem[\protect\citeauthoryear{{Beers}, {Gebhardt}, {Forman}, {Huchra}  \&
  {Jones}}{{Beers} et~al.}{1991}]{Beers_1991}
{Beers} T.~C.,  {Gebhardt} K.,  {Forman} W.,  {Huchra} J.~P.,   {Jones} C.,
  1991, \mn@doi [\aj] {10.1086/115982}, \href
  {https://ui.adsabs.harvard.edu/abs/1991AJ....102.1581B} {102, 1581}

\bibitem[\protect\citeauthoryear{{Bond}, {Cole}, {Efstathiou}  \&
  {Kaiser}}{{Bond} et~al.}{1991}]{Bond_1991}
{Bond} J.~R.,  {Cole} S.,  {Efstathiou} G.,   {Kaiser} N.,  1991, \mn@doi
  [\apj] {10.1086/170520}, \href
  {https://ui.adsabs.harvard.edu/abs/1991ApJ...379..440B} {379, 440}

\bibitem[\protect\citeauthoryear{{Brodwin} et~al.,}{{Brodwin}
  et~al.}{2010}]{Brodwin_2010}
{Brodwin} M.,  et~al., 2010, \mn@doi [\apj] {10.1088/0004-637X/721/1/90}, \href
  {https://ui.adsabs.harvard.edu/\#abs/2010ApJ...721...90B} {721, 90}

\bibitem[\protect\citeauthoryear{{Bruzual} \& {Charlot}}{{Bruzual} \&
  {Charlot}}{2003}]{Bruzual_2003}
{Bruzual} G.,  {Charlot} S.,  2003, \mn@doi [\mnras]
  {10.1046/j.1365-8711.2003.06897.x}, \href
  {https://ui.adsabs.harvard.edu/abs/2003MNRAS.344.1000B} {344, 1000}

\bibitem[\protect\citeauthoryear{{Calzetti}, {Armus}, {Bohlin}, {Kinney},
  {Koornneef}  \& {Storchi-Bergmann}}{{Calzetti} et~al.}{2000}]{Calzetti_2000}
{Calzetti} D.,  {Armus} L.,  {Bohlin} R.~C.,  {Kinney} A.~L.,  {Koornneef} J.,
   {Storchi-Bergmann} T.,  2000, \mn@doi [\apj] {10.1086/308692}, \href
  {https://ui.adsabs.harvard.edu/abs/2000ApJ...533..682C} {533, 682}

\bibitem[\protect\citeauthoryear{{Cassata} et~al.,}{{Cassata}
  et~al.}{2008}]{Cassata_2008}
{Cassata} P.,  et~al., 2008, \mn@doi [\aap] {10.1051/0004-6361:200809881},
  \href {https://ui.adsabs.harvard.edu/abs/2008A&A...483L..39C} {483, L39}

\bibitem[\protect\citeauthoryear{{Chabrier}}{{Chabrier}}{2003}]{Chabrier_2003}
{Chabrier} G.,  2003, \mn@doi [\pasp] {10.1086/376392}, \href
  {https://ui.adsabs.harvard.edu/abs/2003PASP..115..763C} {115, 763}

\bibitem[\protect\citeauthoryear{{Chartab} et~al.,}{{Chartab}
  et~al.}{2019}]{Chartab_2019}
{Chartab} N.,  et~al., 2019, arXiv e-prints, \href
  {https://ui.adsabs.harvard.edu/abs/2019arXiv191204890C} {p. arXiv:1912.04890}

\bibitem[\protect\citeauthoryear{{Cimatti} et~al.,}{{Cimatti}
  et~al.}{2008}]{Cimatti_2008}
{Cimatti} A.,  et~al., 2008, \mn@doi [\aap] {10.1051/0004-6361:20078739}, \href
  {https://ui.adsabs.harvard.edu/abs/2008A&A...482...21C} {482, 21}

\bibitem[\protect\citeauthoryear{{Cooper} et~al.,}{{Cooper}
  et~al.}{2006}]{Cooper_2006}
{Cooper} M.~C.,  et~al., 2006, \mn@doi [\mnras]
  {10.1111/j.1365-2966.2006.10485.x}, \href
  {https://ui.adsabs.harvard.edu/abs/2006MNRAS.370..198C} {370, 198}

\bibitem[\protect\citeauthoryear{{Cooper} et~al.,}{{Cooper}
  et~al.}{2007}]{Cooper_2007}
{Cooper} M.~C.,  et~al., 2007, \mn@doi [\mnras]
  {10.1111/j.1365-2966.2007.11534.x}, \href
  {https://ui.adsabs.harvard.edu/abs/2007MNRAS.376.1445C} {376, 1445}

\bibitem[\protect\citeauthoryear{{Cooper} et~al.,}{{Cooper}
  et~al.}{2008}]{Cooper_2008}
{Cooper} M.~C.,  et~al., 2008, \mn@doi [\mnras]
  {10.1111/j.1365-2966.2007.12613.x}, \href
  {https://ui.adsabs.harvard.edu/abs/2008MNRAS.383.1058C} {383, 1058}

\bibitem[\protect\citeauthoryear{{Davies} et~al.,}{{Davies}
  et~al.}{2016}]{Davies_2016}
{Davies} L.~J.~M.,  et~al., 2016, \mn@doi [\mnras] {10.1093/mnras/stv2573},
  \href {https://ui.adsabs.harvard.edu/abs/2016MNRAS.455.4013D} {455, 4013}

\bibitem[\protect\citeauthoryear{{Demarco} et~al.,}{{Demarco}
  et~al.}{2010}]{Demarco_2010b}
{Demarco} R.,  et~al., 2010, \mn@doi [\apj] {10.1088/0004-637X/711/2/1185},
  \href {https://ui.adsabs.harvard.edu/\#abs/2010ApJ...711.1185D} {711, 1185}

\bibitem[\protect\citeauthoryear{{Duarte} \& {Mamon}}{{Duarte} \&
  {Mamon}}{2015}]{Duarte_2015}
{Duarte} M.,  {Mamon} G.~A.,  2015, \mn@doi [\mnras] {10.1093/mnras/stv1799},
  \href {https://ui.adsabs.harvard.edu/abs/2015MNRAS.453.3848D} {453, 3848}

\bibitem[\protect\citeauthoryear{{Elbaz} et~al.,}{{Elbaz}
  et~al.}{2007}]{Elbaz_2007}
{Elbaz} D.,  et~al., 2007, \mn@doi [\aap] {10.1051/0004-6361:20077525}, \href
  {https://ui.adsabs.harvard.edu/abs/2007A&A...468...33E} {468, 33}

\bibitem[\protect\citeauthoryear{{Erfanianfar} et~al.,}{{Erfanianfar}
  et~al.}{2016}]{Erfanianfar_2016}
{Erfanianfar} G.,  et~al., 2016, \mn@doi [\mnras] {10.1093/mnras/stv2485},
  \href {https://ui.adsabs.harvard.edu/abs/2016MNRAS.455.2839E} {455, 2839}

\bibitem[\protect\citeauthoryear{{Fadda}, {Girardi}, {Giuricin}, {Mardirossian}
   \& {Mezzetti}}{{Fadda} et~al.}{1996}]{Fadda_1996}
{Fadda} D.,  {Girardi} M.,  {Giuricin} G.,  {Mardirossian} F.,   {Mezzetti} M.,
   1996, \mn@doi [\apj] {10.1086/178180}, \href
  {https://ui.adsabs.harvard.edu/abs/1996ApJ...473..670F} {473, 670}

\bibitem[\protect\citeauthoryear{{Fillingham}, {Cooper}, {Wheeler},
  {Garrison-Kimmel}, {Boylan-Kolchin}  \& {Bullock}}{{Fillingham}
  et~al.}{2015}]{Fillingham_2015}
{Fillingham} S.~P.,  {Cooper} M.~C.,  {Wheeler} C.,  {Garrison-Kimmel} S.,
  {Boylan-Kolchin} M.,   {Bullock} J.~S.,  2015, \mn@doi [\mnras]
  {10.1093/mnras/stv2058}, \href
  {https://ui.adsabs.harvard.edu/abs/2015MNRAS.454.2039F} {454, 2039}

\bibitem[\protect\citeauthoryear{{Finoguenov} et~al.,}{{Finoguenov}
  et~al.}{2007}]{Finoguenov_2007}
{Finoguenov} A.,  et~al., 2007, \mn@doi [\apjs] {10.1086/516577}, \href
  {https://ui.adsabs.harvard.edu/abs/2007ApJS..172..182F} {172, 182}

\bibitem[\protect\citeauthoryear{{Finoguenov} et~al.,}{{Finoguenov}
  et~al.}{2010}]{Finoguenov_2010}
{Finoguenov} A.,  et~al., 2010, \mn@doi [\mnras]
  {10.1111/j.1365-2966.2010.16256.x}, \href
  {https://ui.adsabs.harvard.edu/abs/2010MNRAS.403.2063F} {403, 2063}

\bibitem[\protect\citeauthoryear{{Foley} et~al.,}{{Foley}
  et~al.}{2011}]{Foley_2011}
{Foley} R.~J.,  et~al., 2011, \mn@doi [\apj] {10.1088/0004-637X/731/2/86},
  \href {https://ui.adsabs.harvard.edu/\#abs/2011ApJ...731...86F} {731, 86}

\bibitem[\protect\citeauthoryear{{Foltz} et~al.,}{{Foltz}
  et~al.}{2018}]{Foltz_2018}
{Foltz} R.,  et~al., 2018, \mn@doi [\apj] {10.3847/1538-4357/aad80d}, \href
  {https://ui.adsabs.harvard.edu/abs/2018ApJ...866..136F} {866, 136}

\bibitem[\protect\citeauthoryear{{Foreman-Mackey}, {Hogg}, {Lang}  \&
  {Goodman}}{{Foreman-Mackey} et~al.}{2013a}]{Foreman_2013}
{Foreman-Mackey} D.,  {Hogg} D.~W.,  {Lang} D.,   {Goodman} J.,  2013a, \mn@doi
  [\pasp] {10.1086/670067}, \href
  {http://adsabs.harvard.edu/abs/2013PASP..125..306F} {125, 306}

\bibitem[\protect\citeauthoryear{{Foreman-Mackey}, {Hogg}, {Lang}  \&
  {Goodman}}{{Foreman-Mackey} et~al.}{2013b}]{Foreman-Mackey_2013}
{Foreman-Mackey} D.,  {Hogg} D.~W.,  {Lang} D.,   {Goodman} J.,  2013b, \mn@doi
  [\pasp] {10.1086/670067}, \href
  {https://ui.adsabs.harvard.edu/abs/2013PASP..125..306F} {125, 306}

\bibitem[\protect\citeauthoryear{{Fossati} et~al.,}{{Fossati}
  et~al.}{2017}]{Fossati_2017}
{Fossati} M.,  et~al., 2017, \mn@doi [\apj] {10.3847/1538-4357/835/2/153},
  \href {https://ui.adsabs.harvard.edu/abs/2017ApJ...835..153F} {835, 153}

\bibitem[\protect\citeauthoryear{{Galametz} et~al.,}{{Galametz}
  et~al.}{2013}]{Galametz_2013}
{Galametz} A.,  et~al., 2013, \mn@doi [\apjs] {10.1088/0067-0049/206/2/10},
  \href {https://ui.adsabs.harvard.edu/abs/2013ApJS..206...10G} {206, 10}

\bibitem[\protect\citeauthoryear{{Gallazzi} et~al.,}{{Gallazzi}
  et~al.}{2009}]{Gallazzi_2009}
{Gallazzi} A.,  et~al., 2009, \mn@doi [\apj] {10.1088/0004-637X/690/2/1883},
  \href {https://ui.adsabs.harvard.edu/abs/2009ApJ...690.1883G} {690, 1883}

\bibitem[\protect\citeauthoryear{{Gao} \& {White}}{{Gao} \&
  {White}}{2007}]{Gao_2007}
{Gao} L.,  {White} S. D.~M.,  2007, \mn@doi [\mnras]
  {10.1111/j.1745-3933.2007.00292.x}, \href
  {https://ui.adsabs.harvard.edu/abs/2007MNRAS.377L...5G} {377, L5}

\bibitem[\protect\citeauthoryear{{Gao}, {Navarro}, {Cole}, {Frenk}, {White},
  {Springel}, {Jenkins}  \& {Neto}}{{Gao} et~al.}{2008}]{Gao_2008}
{Gao} L.,  {Navarro} J.~F.,  {Cole} S.,  {Frenk} C.~S.,  {White} S. D.~M.,
  {Springel} V.,  {Jenkins} A.,   {Neto} A.~F.,  2008, \mn@doi [\mnras]
  {10.1111/j.1365-2966.2008.13277.x}, \href
  {https://ui.adsabs.harvard.edu/abs/2008MNRAS.387..536G} {387, 536}

\bibitem[\protect\citeauthoryear{{George} et~al.,}{{George}
  et~al.}{2011}]{George_2011}
{George} M.~R.,  et~al., 2011, \mn@doi [\apj] {10.1088/0004-637X/742/2/125},
  \href {https://ui.adsabs.harvard.edu/abs/2011ApJ...742..125G} {742, 125}

\bibitem[\protect\citeauthoryear{{Gilbank}, {Baldry}, {Balogh}, {Glazebrook}
  \& {Bower}}{{Gilbank} et~al.}{2010}]{Gilbank2010}
{Gilbank} D.~G.,  {Baldry} I.~K.,  {Balogh} M.~L.,  {Glazebrook} K.,   {Bower}
  R.~G.,  2010, \mn@doi [\mnras] {10.1111/j.1365-2966.2010.16640.x}, \href
  {https://ui.adsabs.harvard.edu/\#abs/2010MNRAS.405.2594G} {405, 2594}

\bibitem[\protect\citeauthoryear{{Gimeno} et~al.,}{{Gimeno}
  et~al.}{2016}]{Gimeno_2016}
{Gimeno} G.,  et~al., 2016, {On-sky commissioning of Hamamatsu CCDs in GMOS-S}.
p. 99082S, \mn@doi{10.1117/12.2233883}

\bibitem[\protect\citeauthoryear{{Girardi}, {Biviano}, {Giuricin},
  {Mardirossian}  \& {Mezzetti}}{{Girardi} et~al.}{1993}]{Girardi_1993}
{Girardi} M.,  {Biviano} A.,  {Giuricin} G.,  {Mardirossian} F.,   {Mezzetti}
  M.,  1993, \mn@doi [\apj] {10.1086/172256}, \href
  {https://ui.adsabs.harvard.edu/abs/1993ApJ...404...38G} {404, 38}

\bibitem[\protect\citeauthoryear{{Gobat}, {Rosati}, {Strazzullo}, {Rettura},
  {Demarco}  \& {Nonino}}{{Gobat} et~al.}{2008}]{Gobat_2008}
{Gobat} R.,  {Rosati} P.,  {Strazzullo} V.,  {Rettura} A.,  {Demarco} R.,
  {Nonino} M.,  2008, \mn@doi [\aap] {10.1051/0004-6361:200809531}, \href
  {https://ui.adsabs.harvard.edu/abs/2008A&A...488..853G} {488, 853}

\bibitem[\protect\citeauthoryear{{Guglielmo} et~al.,}{{Guglielmo}
  et~al.}{2019}]{Guglielmo_2019}
{Guglielmo} V.,  et~al., 2019, \mn@doi [\aap] {10.1051/0004-6361/201834970},
  \href {https://ui.adsabs.harvard.edu/abs/2019A&A...625A.112G} {625, A112}

\bibitem[\protect\citeauthoryear{{Hahn}, {Tinker}  \& {Wetzel}}{{Hahn}
  et~al.}{2017}]{Hahn_2017}
{Hahn} C.,  {Tinker} J.~L.,   {Wetzel} A.,  2017, \mn@doi [\apj]
  {10.3847/1538-4357/aa6d6b}, \href
  {https://ui.adsabs.harvard.edu/abs/2017ApJ...841....6H} {841, 6}

\bibitem[\protect\citeauthoryear{{Haines} et~al.,}{{Haines}
  et~al.}{2013}]{Haines_2013}
{Haines} C.~P.,  et~al., 2013, \mn@doi [\apj] {10.1088/0004-637X/775/2/126},
  \href {https://ui.adsabs.harvard.edu/abs/2013ApJ...775..126H} {775, 126}

\bibitem[\protect\citeauthoryear{{Hayashi}, {Sobral}, {Best}, {Smail}  \&
  {Kodama}}{{Hayashi} et~al.}{2013}]{Hayashi_2013}
{Hayashi} M.,  {Sobral} D.,  {Best} P.~N.,  {Smail} I.,   {Kodama} T.,  2013,
  \mn@doi [\mnras] {10.1093/mnras/sts676}, \href
  {https://ui.adsabs.harvard.edu/\#abs/2013MNRAS.430.1042H} {430, 1042}

\bibitem[\protect\citeauthoryear{{Hinton}, {Davis}, {Lidman}, {Glazebrook}  \&
  {Lewis}}{{Hinton} et~al.}{2016}]{Hinton_2016}
{Hinton} S.~R.,  {Davis} T.~M.,  {Lidman} C.,  {Glazebrook} K.,   {Lewis}
  G.~F.,  2016, \mn@doi [Astronomy and Computing]
  {10.1016/j.ascom.2016.03.001}, \href
  {https://ui.adsabs.harvard.edu/abs/2016A&C....15...61H} {15, 61}

\bibitem[\protect\citeauthoryear{{Hook}, {J{\o}rgensen}, {Allington-Smith},
  {Davies}, {Metcalfe}, {Murowinski}  \& {Crampton}}{{Hook}
  et~al.}{2004}]{Hook_2004}
{Hook} I.~M.,  {J{\o}rgensen} I.,  {Allington-Smith} J.~R.,  {Davies} R.~L.,
  {Metcalfe} N.,  {Murowinski} R.~G.,   {Crampton} D.,  2004, \mn@doi [\pasp]
  {10.1086/383624}, \href
  {https://ui.adsabs.harvard.edu/abs/2004PASP..116..425H} {116, 425}

\bibitem[\protect\citeauthoryear{Hunter}{Hunter}{2007}]{Hunter_2007}
Hunter J.~D.,  2007, \mn@doi [Computing in Science \& Engineering]
  {10.1109/MCSE.2007.55}, 9, 90

\bibitem[\protect\citeauthoryear{Kass \& Raftery}{Kass \&
  Raftery}{1995}]{kass_1995}
Kass R.~E.,  Raftery A.~E.,  1995, \mn@doi [Journal of the American Statistical
  Association] {10.1080/01621459.1995.10476572}, 90, 773

\bibitem[\protect\citeauthoryear{{Kauffmann} et~al.,}{{Kauffmann}
  et~al.}{2003}]{Kauffmann_2003}
{Kauffmann} G.,  et~al., 2003, \mn@doi [\mnras]
  {10.1111/j.1365-2966.2003.07154.x}, \href
  {https://ui.adsabs.harvard.edu/abs/2003MNRAS.346.1055K} {346, 1055}

\bibitem[\protect\citeauthoryear{{Kauffmann}, {White}, {Heckman}, {M{\'e}nard},
  {Brinchmann}, {Charlot}, {Tremonti}  \& {Brinkmann}}{{Kauffmann}
  et~al.}{2004}]{Kauffmann_2004}
{Kauffmann} G.,  {White} S. D.~M.,  {Heckman} T.~M.,  {M{\'e}nard} B.,
  {Brinchmann} J.,  {Charlot} S.,  {Tremonti} C.,   {Brinkmann} J.,  2004,
  \mn@doi [\mnras] {10.1111/j.1365-2966.2004.08117.x}, \href
  {https://ui.adsabs.harvard.edu/abs/2004MNRAS.353..713K} {353, 713}

\bibitem[\protect\citeauthoryear{{Kausch} et~al.,}{{Kausch}
  et~al.}{2015}]{Kausch_2015}
{Kausch} W.,  et~al., 2015, \mn@doi [\aap] {10.1051/0004-6361/201423909}, \href
  {https://ui.adsabs.harvard.edu/abs/2015A&A...576A..78K} {576, A78}

\bibitem[\protect\citeauthoryear{{Kawinwanichakij} et~al.,}{{Kawinwanichakij}
  et~al.}{2017}]{Kawinwanichakij_2017}
{Kawinwanichakij} L.,  et~al., 2017, \mn@doi [\apj] {10.3847/1538-4357/aa8b75},
  \href {https://ui.adsabs.harvard.edu/abs/2017ApJ...847..134K} {847, 134}

\bibitem[\protect\citeauthoryear{{Kimm} et~al.,}{{Kimm}
  et~al.}{2009}]{Kimm_2009}
{Kimm} T.,  et~al., 2009, \mn@doi [\mnras] {10.1111/j.1365-2966.2009.14414.x},
  \href {https://ui.adsabs.harvard.edu/abs/2009MNRAS.394.1131K} {394, 1131}

\bibitem[\protect\citeauthoryear{{Koyama} et~al.,}{{Koyama}
  et~al.}{2013}]{Koyama_2013}
{Koyama} Y.,  et~al., 2013, \mn@doi [\mnras] {10.1093/mnras/stt1035}, \href
  {https://ui.adsabs.harvard.edu/abs/2013MNRAS.434..423K} {434, 423}

\bibitem[\protect\citeauthoryear{{Kriek}, {van Dokkum}, {Labb{\'e}}, {Franx},
  {Illingworth}, {Marchesini}  \& {Quadri}}{{Kriek} et~al.}{2009}]{Kriek_2009}
{Kriek} M.,  {van Dokkum} P.~G.,  {Labb{\'e}} I.,  {Franx} M.,  {Illingworth}
  G.~D.,  {Marchesini} D.,   {Quadri} R.~F.,  2009, \mn@doi [\apj]
  {10.1088/0004-637X/700/1/221}, \href
  {https://ui.adsabs.harvard.edu/abs/2009ApJ...700..221K} {700, 221}

\bibitem[\protect\citeauthoryear{{Kriek} et~al.,}{{Kriek}
  et~al.}{2018}]{Kriek_2018}
{Kriek} M.,  et~al., 2018, {FAST: Fitting and Assessment of Synthetic
  Templates} (\mn@eprint {ascl} {1803.008})

\bibitem[\protect\citeauthoryear{{Lacey} \& {Cole}}{{Lacey} \&
  {Cole}}{1993}]{Lacey_1993}
{Lacey} C.,  {Cole} S.,  1993, \mn@doi [\mnras] {10.1093/mnras/262.3.627},
  \href {https://ui.adsabs.harvard.edu/abs/1993MNRAS.262..627L} {262, 627}

\bibitem[\protect\citeauthoryear{{Leja}, {Carnall}, {Johnson}, {Conroy}  \&
  {Speagle}}{{Leja} et~al.}{2019a}]{Leja_2019a}
{Leja} J.,  {Carnall} A.~C.,  {Johnson} B.~D.,  {Conroy} C.,   {Speagle} J.~S.,
   2019a, \mn@doi [\apj] {10.3847/1538-4357/ab133c}, \href
  {https://ui.adsabs.harvard.edu/abs/2019ApJ...876....3L} {876, 3}

\bibitem[\protect\citeauthoryear{{Leja} et~al.,}{{Leja}
  et~al.}{2019b}]{Leja_2019b}
{Leja} J.,  et~al., 2019b, \mn@doi [\apj] {10.3847/1538-4357/ab1d5a}, \href
  {https://ui.adsabs.harvard.edu/abs/2019ApJ...877..140L} {877, 140}

\bibitem[\protect\citeauthoryear{{Lotz} et~al.,}{{Lotz}
  et~al.}{2013}]{Lotz_2013}
{Lotz} J.~M.,  et~al., 2013, \mn@doi [\apj] {10.1088/0004-637X/773/2/154},
  \href {https://ui.adsabs.harvard.edu/abs/2013ApJ...773..154L} {773, 154}

\bibitem[\protect\citeauthoryear{{Macci{\`o}}, {Dutton}  \& {van den
  Bosch}}{{Macci{\`o}} et~al.}{2008}]{Maccio_2008}
{Macci{\`o}} A.~V.,  {Dutton} A.~A.,   {van den Bosch} F.~C.,  2008, \mn@doi
  [\mnras] {10.1111/j.1365-2966.2008.14029.x}, \href
  {https://ui.adsabs.harvard.edu/abs/2008MNRAS.391.1940M} {391, 1940}

\bibitem[\protect\citeauthoryear{{Madau} \& {Dickinson}}{{Madau} \&
  {Dickinson}}{2014}]{Madau_2014}
{Madau} P.,  {Dickinson} M.,  2014, \mn@doi [\araa]
  {10.1146/annurev-astro-081811-125615}, \href
  {https://ui.adsabs.harvard.edu/abs/2014ARA&A..52..415M} {52, 415}

\bibitem[\protect\citeauthoryear{{Mamon} \& {Bou{\'e}}}{{Mamon} \&
  {Bou{\'e}}}{2010}]{Mamon_2010}
{Mamon} G.~A.,  {Bou{\'e}} G.,  2010, \mn@doi [\mnras]
  {10.1111/j.1365-2966.2009.15817.x}, \href
  {https://ui.adsabs.harvard.edu/abs/2010MNRAS.401.2433M} {401, 2433}

\bibitem[\protect\citeauthoryear{{Mamon}, {Biviano}  \& {Bou{\'e}}}{{Mamon}
  et~al.}{2013}]{Mamon_2013}
{Mamon} G.~A.,  {Biviano} A.,   {Bou{\'e}} G.,  2013, \mn@doi [\mnras]
  {10.1093/mnras/sts565}, \href
  {https://ui.adsabs.harvard.edu/abs/2013MNRAS.429.3079M} {429, 3079}

\bibitem[\protect\citeauthoryear{{Martini} et~al.,}{{Martini}
  et~al.}{2013}]{Martini_2013}
{Martini} P.,  et~al., 2013, \mn@doi [\apj] {10.1088/0004-637X/768/1/1}, \href
  {https://ui.adsabs.harvard.edu/abs/2013ApJ...768....1M} {768, 1}

\bibitem[\protect\citeauthoryear{{Matharu} et~al.,}{{Matharu}
  et~al.}{2019}]{Matharu_2019}
{Matharu} J.,  et~al., 2019, \mn@doi [\mnras] {10.1093/mnras/sty3465}, \href
  {https://ui.adsabs.harvard.edu/abs/2019MNRAS.484..595M} {484, 595}

\bibitem[\protect\citeauthoryear{{Mauduit} et~al.,}{{Mauduit}
  et~al.}{2012}]{Mauduit_2012}
{Mauduit} J.~C.,  et~al., 2012, \mn@doi [\pasp] {10.1086/666945}, \href
  {https://ui.adsabs.harvard.edu/abs/2012PASP..124..714M} {124, 714}

\bibitem[\protect\citeauthoryear{{McCarthy}, {Frenk}, {Font}, {Lacey}, {Bower},
  {Mitchell}, {Balogh}  \& {Theuns}}{{McCarthy} et~al.}{2008}]{McCarthy_2008}
{McCarthy} I.~G.,  {Frenk} C.~S.,  {Font} A.~S.,  {Lacey} C.~G.,  {Bower}
  R.~G.,  {Mitchell} N.~L.,  {Balogh} M.~L.,   {Theuns} T.,  2008, \mn@doi
  [\mnras] {10.1111/j.1365-2966.2007.12577.x}, \href
  {https://ui.adsabs.harvard.edu/abs/2008MNRAS.383..593M} {383, 593}

\bibitem[\protect\citeauthoryear{{McGee} \& {Balogh}}{{McGee} \&
  {Balogh}}{2010}]{McGee_2010}
{McGee} S.~L.,  {Balogh} M.~L.,  2010, \mn@doi [\mnras]
  {10.1111/j.1365-2966.2010.16616.x}, \href
  {https://ui.adsabs.harvard.edu/abs/2010MNRAS.405.2069M} {405, 2069}

\bibitem[\protect\citeauthoryear{{McGee}, {Bower}  \& {Balogh}}{{McGee}
  et~al.}{2014}]{McGee_2014}
{McGee} S.~L.,  {Bower} R.~G.,   {Balogh} M.~L.,  2014, \mn@doi [\mnras]
  {10.1093/mnrasl/slu066}, \href
  {https://ui.adsabs.harvard.edu/abs/2014MNRAS.442L.105M} {442, L105}

\bibitem[\protect\citeauthoryear{{McLachlan} \& {Basford}}{{McLachlan} \&
  {Basford}}{1988}]{McLachlan_1988}
{McLachlan} G.~J.,  {Basford} K.~E.,  1988, {Mixture models. Inference and
  applications to clustering}.
Statistics: Textbooks and Monographs

\bibitem[\protect\citeauthoryear{{Mok} et~al.,}{{Mok} et~al.}{2013}]{Mok_2013}
{Mok} A.,  et~al., 2013, \mn@doi [\mnras] {10.1093/mnras/stt251}, \href
  {https://ui.adsabs.harvard.edu/abs/2013MNRAS.431.1090M} {431, 1090}

\bibitem[\protect\citeauthoryear{{Mok} et~al.,}{{Mok} et~al.}{2014}]{Mok_2014}
{Mok} A.,  et~al., 2014, \mn@doi [\mnras] {10.1093/mnras/stt2419}, \href
  {https://ui.adsabs.harvard.edu/abs/2014MNRAS.438.3070M} {438, 3070}

\bibitem[\protect\citeauthoryear{{Murowinski} et~al.,}{{Murowinski}
  et~al.}{1998}]{GMOS_1998}
{Murowinski} R.~G.,  et~al., 1998, {Gemini multiobject spectrographs}.
pp 188--195, \mn@doi{10.1117/12.316838}

\bibitem[\protect\citeauthoryear{{Muzzin} et~al.,}{{Muzzin}
  et~al.}{2009}]{Muzzin_2009}
{Muzzin} A.,  et~al., 2009, \mn@doi [\apj] {10.1088/0004-637X/698/2/1934},
  \href {http://adsabs.harvard.edu/abs/2009ApJ...698.1934M} {698, 1934}

\bibitem[\protect\citeauthoryear{{Muzzin} et~al.,}{{Muzzin}
  et~al.}{2012}]{Muzzin_2012}
{Muzzin} A.,  et~al., 2012, \mn@doi [\apj] {10.1088/0004-637X/746/2/188}, \href
  {http://adsabs.harvard.edu/abs/2012ApJ...746..188M} {746, 188}

\bibitem[\protect\citeauthoryear{{Muzzin} et~al.,}{{Muzzin}
  et~al.}{2013}]{Muzzin_2013b}
{Muzzin} A.,  et~al., 2013, \mn@doi [\apjs] {10.1088/0067-0049/206/1/8}, \href
  {https://ui.adsabs.harvard.edu/abs/2013ApJS..206....8M} {206, 8}

\bibitem[\protect\citeauthoryear{{Muzzin} et~al.,}{{Muzzin}
  et~al.}{2014}]{Muzzin_2014}
{Muzzin} A.,  et~al., 2014, \mn@doi [\apj] {10.1088/0004-637X/796/1/65}, \href
  {https://ui.adsabs.harvard.edu/abs/2014ApJ...796...65M} {796, 65}

\bibitem[\protect\citeauthoryear{{Nanayakkara} et~al.,}{{Nanayakkara}
  et~al.}{2016}]{Nanayakkara_2016}
{Nanayakkara} T.,  et~al., 2016, \mn@doi [\apj] {10.3847/0004-637X/828/1/21},
  \href {https://ui.adsabs.harvard.edu/abs/2016ApJ...828...21N} {828, 21}

\bibitem[\protect\citeauthoryear{{Nantais}, {Rettura}, {Lidman}, {Demarco},
  {Gobat}, {Rosati}  \& {Jee}}{{Nantais} et~al.}{2013}]{Nantais_2013b}
{Nantais} J.~B.,  {Rettura} A.,  {Lidman} C.,  {Demarco} R.,  {Gobat} R.,
  {Rosati} P.,   {Jee} M.~J.,  2013, \mn@doi [\aap]
  {10.1051/0004-6361/201321877}, \href
  {https://ui.adsabs.harvard.edu/abs/2013A&A...556A.112N} {556, A112}

\bibitem[\protect\citeauthoryear{{Nantais} et~al.,}{{Nantais}
  et~al.}{2017}]{Nantais_2017}
{Nantais} J.~B.,  et~al., 2017, \mn@doi [\mnras] {10.1093/mnrasl/slw224}, \href
  {https://ui.adsabs.harvard.edu/abs/2017MNRAS.465L.104N} {465, L104}

\bibitem[\protect\citeauthoryear{{Navarro}, {Frenk}  \& {White}}{{Navarro}
  et~al.}{1997}]{Navarro_1997}
{Navarro} J.~F.,  {Frenk} C.~S.,   {White} S. D.~M.,  1997, \mn@doi [\apj]
  {10.1086/304888}, \href
  {https://ui.adsabs.harvard.edu/abs/1997ApJ...490..493N} {490, 493}

\bibitem[\protect\citeauthoryear{{Neistein} \& {Dekel}}{{Neistein} \&
  {Dekel}}{2008}]{Neistein_2008}
{Neistein} E.,  {Dekel} A.,  2008, \mn@doi [\mnras]
  {10.1111/j.1365-2966.2007.12570.x}, \href
  {https://ui.adsabs.harvard.edu/abs/2008MNRAS.383..615N} {383, 615}

\bibitem[\protect\citeauthoryear{{Neistein}, {van den Bosch}  \&
  {Dekel}}{{Neistein} et~al.}{2006}]{Neistein_2006}
{Neistein} E.,  {van den Bosch} F.~C.,   {Dekel} A.,  2006, \mn@doi [\mnras]
  {10.1111/j.1365-2966.2006.10918.x}, \href
  {https://ui.adsabs.harvard.edu/abs/2006MNRAS.372..933N} {372, 933}

\bibitem[\protect\citeauthoryear{{Newman}, {Ellis}, {Andreon}, {Treu},
  {Raichoor}  \& {Trinchieri}}{{Newman} et~al.}{2014}]{Newman_2014}
{Newman} A.~B.,  {Ellis} R.~S.,  {Andreon} S.,  {Treu} T.,  {Raichoor} A.,
  {Trinchieri} G.,  2014, \mn@doi [\apj] {10.1088/0004-637X/788/1/51}, \href
  {https://ui.adsabs.harvard.edu/abs/2014ApJ...788...51N} {788, 51}

\bibitem[\protect\citeauthoryear{{Noeske} et~al.,}{{Noeske}
  et~al.}{2007}]{Noeske_2007}
{Noeske} K.~G.,  et~al., 2007, \mn@doi [\apjl] {10.1086/517926}, \href
  {https://ui.adsabs.harvard.edu/abs/2007ApJ...660L..43N} {660, L43}

\bibitem[\protect\citeauthoryear{{Noirot} et~al.,}{{Noirot}
  et~al.}{2018}]{Noirot_2018}
{Noirot} G.,  et~al., 2018, \mn@doi [\apj] {10.3847/1538-4357/aabadb}, \href
  {https://ui.adsabs.harvard.edu/abs/2018ApJ...859...38N} {859, 38}

\bibitem[\protect\citeauthoryear{Oliphant}{Oliphant}{2015}]{Oliphant_2015}
Oliphant T.~E.,  2015, Guide to NumPy, 2nd edn.
CreateSpace Independent Publishing Platform, USA

\bibitem[\protect\citeauthoryear{{Oman} \& {Hudson}}{{Oman} \&
  {Hudson}}{2016}]{Oman_2016}
{Oman} K.~A.,  {Hudson} M.~J.,  2016, \mn@doi [\mnras] {10.1093/mnras/stw2195},
  \href {https://ui.adsabs.harvard.edu/abs/2016MNRAS.463.3083O} {463, 3083}

\bibitem[\protect\citeauthoryear{{Paccagnella} et~al.,}{{Paccagnella}
  et~al.}{2016}]{Paccagnella_2016}
{Paccagnella} A.,  et~al., 2016, \mn@doi [\apjl] {10.3847/2041-8205/816/2/L25},
  \href {https://ui.adsabs.harvard.edu/abs/2016ApJ...816L..25P} {816, L25}

\bibitem[\protect\citeauthoryear{{Patel}, {Kelson}, {Holden}, {Franx}  \&
  {Illingworth}}{{Patel} et~al.}{2011}]{Patel_2011}
{Patel} S.~G.,  {Kelson} D.~D.,  {Holden} B.~P.,  {Franx} M.,   {Illingworth}
  G.~D.,  2011, \mn@doi [\apj] {10.1088/0004-637X/735/1/53}, \href
  {https://ui.adsabs.harvard.edu/abs/2011ApJ...735...53P} {735, 53}

\bibitem[\protect\citeauthoryear{{Peng} et~al.,}{{Peng}
  et~al.}{2010}]{Peng_2010}
{Peng} Y.-j.,  et~al., 2010, \mn@doi [\apj] {10.1088/0004-637X/721/1/193},
  \href {https://ui.adsabs.harvard.edu/abs/2010ApJ...721..193P} {721, 193}

\bibitem[\protect\citeauthoryear{{Pintos-Castro}, {Yee}, {Muzzin}, {Old}  \&
  {Wilson}}{{Pintos-Castro} et~al.}{2019}]{Pintos_Castro_2019}
{Pintos-Castro} I.,  {Yee} H.~K.~C.,  {Muzzin} A.,  {Old} L.,   {Wilson} G.,
  2019, \mn@doi [\apj] {10.3847/1538-4357/ab14ee}, \href
  {https://ui.adsabs.harvard.edu/abs/2019ApJ...876...40P} {876, 40}

\bibitem[\protect\citeauthoryear{{Poggianti} et~al.,}{{Poggianti}
  et~al.}{2006}]{Poggianti_2006}
{Poggianti} B.~M.,  et~al., 2006, \mn@doi [\apj] {10.1086/500666}, \href
  {https://ui.adsabs.harvard.edu/abs/2006ApJ...642..188P} {642, 188}

\bibitem[\protect\citeauthoryear{{Popesso} et~al.,}{{Popesso}
  et~al.}{2011}]{Popesso_2011}
{Popesso} P.,  et~al., 2011, \mn@doi [\aap] {10.1051/0004-6361/201015672},
  \href {https://ui.adsabs.harvard.edu/abs/2011A&A...532A.145P} {532, A145}

\bibitem[\protect\citeauthoryear{{Press} \& {Schechter}}{{Press} \&
  {Schechter}}{1974}]{Press_1974}
{Press} W.~H.,  {Schechter} P.,  1974, \mn@doi [\apj] {10.1086/152650}, \href
  {https://ui.adsabs.harvard.edu/abs/1974ApJ...187..425P} {187, 425}

\bibitem[\protect\citeauthoryear{{Rodr{\'\i}guez del Pino}
  et~al.,}{{Rodr{\'\i}guez del Pino} et~al.}{2017}]{Rodriguez_2017}
{Rodr{\'\i}guez del Pino} B.,  et~al., 2017, \mn@doi [\mnras]
  {10.1093/mnras/stx228}, \href
  {https://ui.adsabs.harvard.edu/abs/2017MNRAS.467.4200R} {467, 4200}

\bibitem[\protect\citeauthoryear{{Saglia} et~al.,}{{Saglia}
  et~al.}{2010}]{Saglia_2010}
{Saglia} R.~P.,  et~al., 2010, \mn@doi [\aap] {10.1051/0004-6361/201014703},
  \href {https://ui.adsabs.harvard.edu/abs/2010A&A...524A...6S} {524, A6}

\bibitem[\protect\citeauthoryear{{Sanders} et~al.,}{{Sanders}
  et~al.}{2007}]{Sanders_2007}
{Sanders} D.~B.,  et~al., 2007, \mn@doi [\apjs] {10.1086/517885}, \href
  {https://ui.adsabs.harvard.edu/abs/2007ApJS..172...86S} {172, 86}

\bibitem[\protect\citeauthoryear{{Schreiber} et~al.,}{{Schreiber}
  et~al.}{2015}]{Schreiber_2015}
{Schreiber} C.,  et~al., 2015, \mn@doi [\aap] {10.1051/0004-6361/201425017},
  \href {https://ui.adsabs.harvard.edu/abs/2015A%26A...575A..74S} {575, A74}

\bibitem[\protect\citeauthoryear{Schwarz}{Schwarz}{1978}]{Schwarz_1978}
Schwarz G.,  1978, Annals Statist., 6, 461

\bibitem[\protect\citeauthoryear{{Sen}}{{Sen}}{1968}]{Sen_1968}
{Sen} P.~K.,  1968, \mn@doi [Journal of the American Statistical Association]
  {10.2307/2285891}, 63, 1379–1389

\bibitem[\protect\citeauthoryear{{Smette} et~al.,}{{Smette}
  et~al.}{2015}]{Smette_2015}
{Smette} A.,  et~al., 2015, \mn@doi [\aap] {10.1051/0004-6361/201423932}, \href
  {https://ui.adsabs.harvard.edu/abs/2015A&A...576A..77S} {576, A77}

\bibitem[\protect\citeauthoryear{{Snyder} et~al.,}{{Snyder}
  et~al.}{2012}]{Snyder_2012}
{Snyder} G.~F.,  et~al., 2012, \mn@doi [\apj] {10.1088/0004-637X/756/2/114},
  \href {https://ui.adsabs.harvard.edu/abs/2012ApJ...756..114S} {756, 114}

\bibitem[\protect\citeauthoryear{{Sobral}, {Best}, {Matsuda}, {Smail}, {Geach}
  \& {Cirasuolo}}{{Sobral} et~al.}{2012}]{Sobral_2012}
{Sobral} D.,  {Best} P.~N.,  {Matsuda} Y.,  {Smail} I.,  {Geach} J.~E.,
  {Cirasuolo} M.,  2012, \mn@doi [\mnras] {10.1111/j.1365-2966.2011.19977.x},
  \href {http://adsabs.harvard.edu/abs/2012MNRAS.420.1926S} {420, 1926}

\bibitem[\protect\citeauthoryear{{Stalder} et~al.,}{{Stalder}
  et~al.}{2013}]{Stalder_2013}
{Stalder} B.,  et~al., 2013, \mn@doi [\apj] {10.1088/0004-637X/763/2/93}, \href
  {https://ui.adsabs.harvard.edu/\#abs/2013ApJ...763...93S} {763, 93}

\bibitem[\protect\citeauthoryear{{Stanford}, {Gonzalez}, {Brodwin}, {Gettings},
  {Eisenhardt}, {Stern}  \& {Wylezalek}}{{Stanford}
  et~al.}{2014}]{Stanford_2014}
{Stanford} S.~A.,  {Gonzalez} A.~H.,  {Brodwin} M.,  {Gettings} D.~P.,
  {Eisenhardt} P.~R.~M.,  {Stern} D.,   {Wylezalek} D.,  2014, VizieR Online
  Data Catalog, \href {https://ui.adsabs.harvard.edu/abs/2014yCat..22130025S}
  {221}

\bibitem[\protect\citeauthoryear{{Strateva} et~al.,}{{Strateva}
  et~al.}{2001}]{Strateva_2001}
{Strateva} I.,  et~al., 2001, \mn@doi [\aj] {10.1086/323301}, \href
  {https://ui.adsabs.harvard.edu/abs/2001AJ....122.1861S} {122, 1861}

\bibitem[\protect\citeauthoryear{{Strazzullo} et~al.,}{{Strazzullo}
  et~al.}{2006}]{Strazzullo_2006}
{Strazzullo} V.,  et~al., 2006, \mn@doi [\aap] {10.1051/0004-6361:20054341},
  \href {https://ui.adsabs.harvard.edu/abs/2006A&A...450..909S} {450, 909}

\bibitem[\protect\citeauthoryear{{Strazzullo} et~al.,}{{Strazzullo}
  et~al.}{2019}]{Strazzullo_2019}
{Strazzullo} V.,  et~al., 2019, \mn@doi [\aap] {10.1051/0004-6361/201833944},
  \href {https://ui.adsabs.harvard.edu/abs/2019A&A...622A.117S} {622, A117}

\bibitem[\protect\citeauthoryear{{Tacconi} et~al.,}{{Tacconi}
  et~al.}{2013}]{Tacconi_2013}
{Tacconi} L.~J.,  et~al., 2013, \mn@doi [\apj] {10.1088/0004-637X/768/1/74},
  \href {https://ui.adsabs.harvard.edu/abs/2013ApJ...768...74T} {768, 74}

\bibitem[\protect\citeauthoryear{{Tacconi} et~al.,}{{Tacconi}
  et~al.}{2018}]{Tacconi_2018}
{Tacconi} L.~J.,  et~al., 2018, \mn@doi [\apj] {10.3847/1538-4357/aaa4b4},
  \href {https://ui.adsabs.harvard.edu/abs/2018ApJ...853..179T} {853, 179}

\bibitem[\protect\citeauthoryear{{Taranu}, {Hudson}, {Balogh}, {Smith},
  {Power}, {Oman}  \& {Krane}}{{Taranu} et~al.}{2014}]{Taranu_2014}
{Taranu} D.~S.,  {Hudson} M.~J.,  {Balogh} M.~L.,  {Smith} R.~J.,  {Power} C.,
  {Oman} K.~A.,   {Krane} B.,  2014, \mn@doi [\mnras] {10.1093/mnras/stu389},
  \href {https://ui.adsabs.harvard.edu/abs/2014MNRAS.440.1934T} {440, 1934}

\bibitem[\protect\citeauthoryear{{Taylor} et~al.,}{{Taylor}
  et~al.}{2015}]{Taylor_2015}
{Taylor} E.~N.,  et~al., 2015, \mn@doi [\mnras] {10.1093/mnras/stu1900}, \href
  {https://ui.adsabs.harvard.edu/abs/2015MNRAS.446.2144T} {446, 2144}

\bibitem[\protect\citeauthoryear{{Theil}}{{Theil}}{1950}]{Theil_1950}
{Theil} H.,  1950, in Proceedings of Koninklijke Nederlandse Akademie
  Wetenschappen, Series A Mathematical Sciences. pp 386–392,85–91

\bibitem[\protect\citeauthoryear{{Tinker} \& {Wetzel}}{{Tinker} \&
  {Wetzel}}{2010}]{Tinker_2010}
{Tinker} J.~L.,  {Wetzel} A.~R.,  2010, \mn@doi [\apj]
  {10.1088/0004-637X/719/1/88}, \href
  {https://ui.adsabs.harvard.edu/abs/2010ApJ...719...88T} {719, 88}

\bibitem[\protect\citeauthoryear{{Tinker}, {Leauthaud}, {Bundy}, {George},
  {Behroozi}, {Massey}, {Rhodes}  \& {Wechsler}}{{Tinker}
  et~al.}{2013}]{Tinker_2013}
{Tinker} J.~L.,  {Leauthaud} A.,  {Bundy} K.,  {George} M.~R.,  {Behroozi} P.,
  {Massey} R.,  {Rhodes} J.,   {Wechsler} R.~H.,  2013, \mn@doi [\apj]
  {10.1088/0004-637X/778/2/93}, \href
  {https://ui.adsabs.harvard.edu/abs/2013ApJ...778...93T} {778, 93}

\bibitem[\protect\citeauthoryear{{Virtanen} et~al.,}{{Virtanen}
  et~al.}{2019}]{SciPy_2019}
{Virtanen} P.,  et~al., 2019, arXiv e-prints, \href
  {https://ui.adsabs.harvard.edu/abs/2019arXiv190710121V} {p. arXiv:1907.10121}

\bibitem[\protect\citeauthoryear{{Vulcani}, {Poggianti}, {Finn}, {Rudnick},
  {Desai}  \& {Bamford}}{{Vulcani} et~al.}{2010}]{Vulcani_2010}
{Vulcani} B.,  {Poggianti} B.~M.,  {Finn} R.~A.,  {Rudnick} G.,  {Desai} V.,
  {Bamford} S.,  2010, \mn@doi [\apjl] {10.1088/2041-8205/710/1/L1}, \href
  {https://ui.adsabs.harvard.edu/abs/2010ApJ...710L...1V} {710, L1}

\bibitem[\protect\citeauthoryear{{Wang} et~al.,}{{Wang}
  et~al.}{2018}]{Wang_2018}
{Wang} L.,  et~al., 2018, \mn@doi [\aap] {10.1051/0004-6361/201832697}, \href
  {https://ui.adsabs.harvard.edu/abs/2018A&A...618A...1W} {618, A1}

\bibitem[\protect\citeauthoryear{{Wechsler}, {Zentner}, {Bullock}, {Kravtsov}
  \& {Allgood}}{{Wechsler} et~al.}{2006}]{Wechsler_2006}
{Wechsler} R.~H.,  {Zentner} A.~R.,  {Bullock} J.~S.,  {Kravtsov} A.~V.,
  {Allgood} B.,  2006, \mn@doi [\apj] {10.1086/507120}, \href
  {https://ui.adsabs.harvard.edu/abs/2006ApJ...652...71W} {652, 71}

\bibitem[\protect\citeauthoryear{{Wetzel}, {Tinker}  \& {Conroy}}{{Wetzel}
  et~al.}{2012}]{Wetzel_2012}
{Wetzel} A.~R.,  {Tinker} J.~L.,   {Conroy} C.,  2012, \mn@doi [\mnras]
  {10.1111/j.1365-2966.2012.21188.x}, \href
  {https://ui.adsabs.harvard.edu/abs/2012MNRAS.424..232W} {424, 232}

\bibitem[\protect\citeauthoryear{{Wetzel}, {Tinker}, {Conroy}  \& {van den
  Bosch}}{{Wetzel} et~al.}{2013}]{Wetzel_2013}
{Wetzel} A.~R.,  {Tinker} J.~L.,  {Conroy} C.,   {van den Bosch} F.~C.,  2013,
  \mn@doi [\mnras] {10.1093/mnras/stt469}, \href
  {https://ui.adsabs.harvard.edu/abs/2013MNRAS.432..336W} {432, 336}

\bibitem[\protect\citeauthoryear{{Whitaker}, {van Dokkum}, {Brammer}  \&
  {Franx}}{{Whitaker} et~al.}{2012}]{Whitaker_2012}
{Whitaker} K.~E.,  {van Dokkum} P.~G.,  {Brammer} G.,   {Franx} M.,  2012,
  \mn@doi [\apjl] {10.1088/2041-8205/754/2/L29}, \href
  {https://ui.adsabs.harvard.edu/abs/2012ApJ...754L..29W} {754, L29}

\bibitem[\protect\citeauthoryear{{Whitaker} et~al.,}{{Whitaker}
  et~al.}{2014}]{Whitaker_2014}
{Whitaker} K.~E.,  et~al., 2014, \mn@doi [\apj] {10.1088/0004-637X/795/2/104},
  \href {https://ui.adsabs.harvard.edu/abs/2014ApJ...795..104W} {795, 104}

\bibitem[\protect\citeauthoryear{{Wijesinghe} et~al.,}{{Wijesinghe}
  et~al.}{2012}]{Wijesinghe_2012}
{Wijesinghe} D.~B.,  et~al., 2012, \mn@doi [\mnras]
  {10.1111/j.1365-2966.2012.21164.x}, \href
  {https://ui.adsabs.harvard.edu/abs/2012MNRAS.423.3679W} {423, 3679}

\bibitem[\protect\citeauthoryear{{Williams}, {Quadri}, {Franx}, {van Dokkum}
  \& {Labb{\'e}}}{{Williams} et~al.}{2009}]{Williams_2009}
{Williams} R.~J.,  {Quadri} R.~F.,  {Franx} M.,  {van Dokkum} P.,   {Labb{\'e}}
  I.,  2009, \mn@doi [\apj] {10.1088/0004-637X/691/2/1879}, \href
  {https://ui.adsabs.harvard.edu/abs/2009ApJ...691.1879W} {691, 1879}

\bibitem[\protect\citeauthoryear{{Wilson} et~al.,}{{Wilson}
  et~al.}{2009}]{Wilson_2009}
{Wilson} G.,  et~al., 2009, \mn@doi [\apj] {10.1088/0004-637X/698/2/1943},
  \href {http://adsabs.harvard.edu/abs/2009ApJ...698.1943W} {698, 1943}

\bibitem[\protect\citeauthoryear{{Wojtak} et~al.,}{{Wojtak}
  et~al.}{2018}]{Wojtak_2018}
{Wojtak} R.,  et~al., 2018, \mn@doi [\mnras] {10.1093/mnras/sty2257}, \href
  {https://ui.adsabs.harvard.edu/abs/2018MNRAS.481..324W} {481, 324}

\bibitem[\protect\citeauthoryear{{Yang}, {Mo}, {van den Bosch}  \&
  {Jing}}{{Yang} et~al.}{2005}]{Yang_2005}
{Yang} X.,  {Mo} H.~J.,  {van den Bosch} F.~C.,   {Jing} Y.~P.,  2005, \mn@doi
  [\mnras] {10.1111/j.1365-2966.2005.08560.x}, \href
  {https://ui.adsabs.harvard.edu/abs/2005MNRAS.356.1293Y} {356, 1293}

\bibitem[\protect\citeauthoryear{{York} et~al.,}{{York}
  et~al.}{2000}]{York_2000}
{York} D.~G.,  et~al., 2000, \mn@doi [\aj] {10.1086/301513}, \href
  {https://ui.adsabs.harvard.edu/abs/2000AJ....120.1579Y} {120, 1579}

\bibitem[\protect\citeauthoryear{{Zeimann} et~al.,}{{Zeimann}
  et~al.}{2013}]{Zeimann_2013}
{Zeimann} G.~R.,  et~al., 2013, \mn@doi [\apj] {10.1088/0004-637X/779/2/137},
  \href {https://ui.adsabs.harvard.edu/abs/2013ApJ...779..137Z} {779, 137}

\bibitem[\protect\citeauthoryear{{Zentner}, {Hearin}  \& {van den
  Bosch}}{{Zentner} et~al.}{2014}]{Zentner_2014}
{Zentner} A.~R.,  {Hearin} A.~P.,   {van den Bosch} F.~C.,  2014, \mn@doi
  [\mnras] {10.1093/mnras/stu1383}, \href
  {https://ui.adsabs.harvard.edu/abs/2014MNRAS.443.3044Z} {443, 3044}

\bibitem[\protect\citeauthoryear{{van Dokkum} \& {van der Marel}}{{van Dokkum}
  \& {van der Marel}}{2007}]{van_Dokkum_2007}
{van Dokkum} P.~G.,  {van der Marel} R.~P.,  2007, \mn@doi [\apj]
  {10.1086/509633}, \href
  {https://ui.adsabs.harvard.edu/abs/2007ApJ...655...30V} {655, 30}

\bibitem[\protect\citeauthoryear{{von der Linden}, {Wild}, {Kauffmann}, {White}
   \& {Weinmann}}{{von der Linden} et~al.}{2010}]{vonderlinden_2010}
{von der Linden} A.,  {Wild} V.,  {Kauffmann} G.,  {White} S.~D.~M.,
  {Weinmann} S.,  2010, \mn@doi [\mnras] {10.1111/j.1365-2966.2010.16375.x},
  \href {https://ui.adsabs.harvard.edu/abs/2010MNRAS.404.1231V} {404, 1231}

\makeatother
\end{thebibliography}


\begin{thebibliography}{}
\makeatletter
\relax
\def\mn@urlcharsother{\let\do\@makeother \do\$\do\&\do\#\do\^\do\_\do\%\do\~}
\def\mn@doi{\begingroup\mn@urlcharsother \@ifnextchar [ {\mn@doi@}
  {\mn@doi@[]}}
\def\mn@doi@[#1]#2{\def\@tempa{#1}\ifx\@tempa\@empty \href
  {http://dx.doi.org/#2} {doi:#2}\else \href {http://dx.doi.org/#2} {#1}\fi
  \endgroup}
\def\mn@eprint#1#2{\mn@eprint@#1:#2::\@nil}
\def\mn@eprint@arXiv#1{\href {http://arxiv.org/abs/#1} {{\tt arXiv:#1}}}
\def\mn@eprint@dblp#1{\href {http://dblp.uni-trier.de/rec/bibtex/#1.xml}
  {dblp:#1}}
\def\mn@eprint@#1:#2:#3:#4\@nil{\def\@tempa {#1}\def\@tempb {#2}\def\@tempc
  {#3}\ifx \@tempc \@empty \let \@tempc \@tempb \let \@tempb \@tempa \fi \ifx
  \@tempb \@empty \def\@tempb {arXiv}\fi \@ifundefined
  {mn@eprint@\@tempb}{\@tempb:\@tempc}{\expandafter \expandafter \csname
  mn@eprint@\@tempb\endcsname \expandafter{\@tempc}}}

\bibitem[\protect\citeauthoryear{{Muzzin} et~al.,}{{Muzzin}
  et~al.}{2013}]{Muzzin_2013b}
{Muzzin} A.,  et~al., 2013, \mn@doi [\apjs] {10.1088/0067-0049/206/1/8}, \href
  {https://ui.adsabs.harvard.edu/abs/2013ApJS..206....8M} {206, 8}

\bibitem[\protect\citeauthoryear{{Schreiber} et~al.,}{{Schreiber}
  et~al.}{2015}]{Schreiber_2015}
{Schreiber} C.,  et~al., 2015, \mn@doi [\aap] {10.1051/0004-6361/201425017},
  \href {https://ui.adsabs.harvard.edu/abs/2015A%26A...575A..74S} {575, A74}

\bibitem[\protect\citeauthoryear{{Williams}, {Quadri}, {Franx}, {van Dokkum}
  \& {Labb{\'e}}}{{Williams} et~al.}{2009}]{Williams_2009}
{Williams} R.~J.,  {Quadri} R.~F.,  {Franx} M.,  {van Dokkum} P.,   {Labb{\'e}}
  I.,  2009, \mn@doi [\apj] {10.1088/0004-637X/691/2/1879}, \href
  {https://ui.adsabs.harvard.edu/abs/2009ApJ...691.1879W} {691, 1879}

\bibitem[\protect\citeauthoryear{{van der Burg} et~al.,}{{van der Burg}
  et~al.}{2020}]{vdburg_2020}
{van der Burg} R. F.~J.,  et~al., 2020, \mn@doi [\aap]
  {10.1051/0004-6361/202037754}, \href
  {https://ui.adsabs.harvard.edu/abs/2020A&A...638A.112V} {638, A112}

\makeatother
\end{thebibliography}


\section*{Affiliations}
$^{1}$European Space Agency (ESA), European Space Astronomy Centre, Villanueva de la Ca\~{n}ada, E-28691 Madrid, Spain\\
$^{2}$Department of Astronomy $\&$ Astrophysics, University of Toronto, Toronto, Canada\\
$^{3}$Department of Physics and Astronomy, University of Waterloo, Waterloo, Ontario N2L 3G1, Canada\\
$^{4}$European Southern Observatory, Karl-Schwarzschild-Str. 2, 85748, Garching, Germany\\
$^{5}$INAF – Osservatorio Astronomico di Trieste, via G. B. Tiepolo 11, I-34143
Trieste, Italy\\
$^{6}$IFPU - Institute for Fundamental Physics of the Universe, via Beirut 2, 34014 Trieste, Italy\\
$^{7}$Department of Physics and Astronomy, York University, 4700 Keele Street,
Toronto, Ontario, ON MJ3 1P3, Canada\\
$^{8}$Department of Physics and Astronomy, The University of Kansas, 1251
Wescoe Hall Drive, Lawrence, KS 66045, USA\\
$^{9}$INAF - Osservatorio astronomico di Padova, Vicolo Osservatorio 5, IT-35122 Padova, Italy\\
$^{10}$Department of Physics and Astronomy, University of California, Irvine,
4129 Frederick Reines Hall, Irvine, CA 92697, USA\\
$^{11}$Steward Observatory and Department of Astronomy, University of
Arizona, Tucson, AZ 85719, USA\\
$^{12}$Departamento de Astronom\'ia, Facultad de Ciencias F\'isicas y Matem\'aticas, Universidad de Concepci\'on, Concepci\'on, Chile\\
$^{13}$Department of Physics and Astronomy, University of California, Riverside,
900 University Avenue, Riverside, CA 92521, USA\\
$^{14}$Australian Astronomical Observatory, 105 Delhi Road, North Ryde, NSW
2113, Australia\\
$^{15}$School of Physics and Astronomy, University of Birmingham, Edgbaston,
Birmingham B15 2TT, England\\
$^{16}$South African Astronomical Observatory, P.O. Box 9, Observatory 7935
Cape Town, South Africa\\
$^{17}$Centre for Space Research, North-West University, Potchefstroom 2520
Cape Town, South Africa\\
$^{18}$Astrophysics Research Institute, Liverpool John Moores University, 146 Brownlow Hill, Liverpool L3 5RF, UK\\
$^{19}$Laboratoire d'astrophysique, \'Ecole
Polytechnique F\'ed\'erale de Lausanne,
Switzerland\\
$^{20}$Departamento de Ciencias F\'{i}sicas, Universidad Andres Bello, Fernandez Concha 700, Las Condes 7591538, Santiago, Regi\'{o}n Metropolitana, Chile\\
$^{21}$Arizona State University, School of Earth and Space Exploration, Tempe, AZ 871404, USA\\
$^{22}$MIT Kavli Institute for Astrophysics and Space Research, 70 Vassar St,
Cambridge, MA 02109, USA\\
$^{23}$Department of Physics, McGill University, 3600 rue University, Montr\'{e}al, Qu\'{e}bec, H3P 1T3, Canada

\appendix

\section{Cluster membership algorithms}
\label{sec:appendix_cluster_membership}
 \begin{figure*}
	\includegraphics[width=1.0\textwidth]{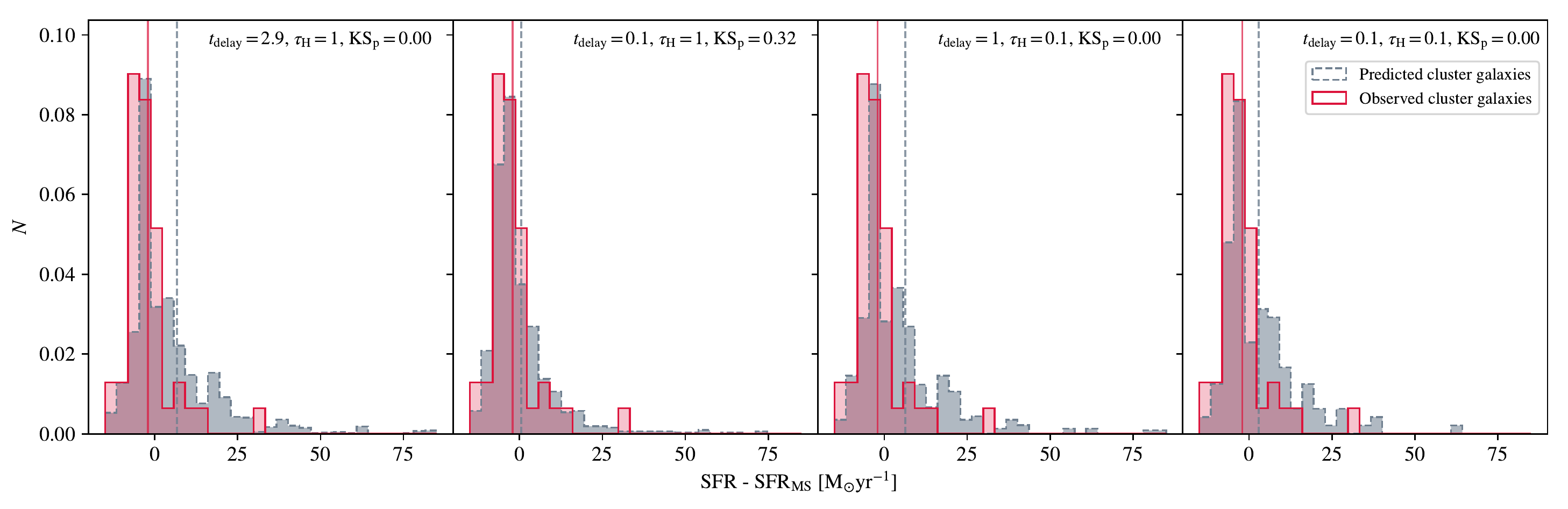}
    \vspace*{-5mm}\caption{Example toy model-predicted cluster and observed cluster $\rm \Delta SFR_{\rm MS}$ distributions for different $
    \tau_{\rm H}$ and $t_{\rm delay}$ timescales.}
    \label{fig:model_delta_MS_SFR_hist_specific_examples_i_4_1.0_z_1.3_SFR_zlim_1.0}
\end{figure*}
In this Section, we describe the two algorithms used to define cluster membership. First, the main cluster redshift peak is identified by selecting those galaxies in the cluster field with $c |z-z_{\rm c}|\leq \rm{6000 kms^{-1}}$ (as in \citealt{Beers_1991,Girardi_1993}). Here, $c$ is the speed of light and $z_{\rm c}$ is the  cluster redshift estimate from \citet{Balogh_2017}. To try to identify cases of merging subclusters close to the line-of-sight, and when these cases occur, to separate these subcluster components from the main cluster, the KMM algorithm is applied to the distribution of redshifts located in the main peak. The algorithm estimates the probability that the $z$ distribution is better represented by $k$ Gaussians rather than a single Gaussian \citep{McLachlan_1988,Ashman_1994}.

With the resulting galaxies left after the main-peak and KMM selection procedures, cluster membership is then refined using two techniques, Clean (\citealt{Mamon_2013}), and C.L.U.M.P.S (Munari et al. in preparation). Both these algorithms identify cluster members based on their location in projected phase-space, $R\nu_{\rm rf}$, where $R$ is the projected radial distance from the cluster center, and $\nu_{\rm rf}\equiv c(z-\bar{z})/(1+\bar{z})$ is the rest-frame velocity. While these two algorithms are both based in projected phase-space, they are conceptually very different.

The Clean algorithm is theoretically motivated, with its parameters fixed by properties of cluster-sized haloes extracted from cosmological numerical simulations. The Clean method uses an estimate of the cluster line-of-sight velocity dispersion $\sigma_{\rm los}$, to predict the cluster mass from a scaling relation. The algorithm then adopts a NFW profile (\citealt{Navarro_1997}), a theoretical concentration-mass relation (\citealt{Maccio_2008}), and a velocity anisotropy profile model (\citealt{Mamon_2010}), to predict $\sigma_{\rm los}(R)$, and to iteratively reject galaxies with $|\nu_{\rm rf}|>2.7\sigma_{\rm los}$ at any radius. 

The C.L.U.M.P.S algorithm is based only on the fact that clusters of galaxies manifest themselves as concentrations in project phase-space, and so by nature, it is less model-dependent than the Clean algorithm. The C.L.U.M.P.S is based on the Shifting Gapper (SG) method of \citet{Fadda_1996}, however is more robust to the SG method with respect to the choice of the initial parameters that define the smoothing lengths in projected phase-space.  The C.L.U.M.P.S method evaluates the density of galaxies in projected phase-space, and convolves this density map with a Gaussian filter in Fourier space to remove high frequencies. The technique then bins this smoothed density along the radial direction to identify the main peak in velocity space. The minima of this peak define the velocity limits within which to include cluster members in that given radial bin. We note that the method is still being refined and tested (Munari et al. in preparation).

These two algorithms are applied to the data twice, the first time where $\bar{z}$ is defined as the average redshift of the galaxies that were selected during the main-peak and KMM procedures, and the second time where $\bar{z}$ is defined as the average redshift of the galaxies selected as members from the first run. The radius, $r_{\rm 200c}$ is obtained from $\sigma_{\rm los}$ and equation B3 in \citet{Mamon_2013} in an iterative procedure where we assume the \citet{Mamon_2010} velocity anisotropy profile, and a NFW profile model for the mass distribution with a concentration taken to be $c_{\rm 200}=5$ on the first iteration, and derived from the concentration-mass relation of \citet{Gao_2008} on subsequent iterations. In this paper, cluster members are defined as those that are identified by either the Clean or C.L.U.M.P.S algorithm. 

The number of galaxies identified by the Clean algorithm as cluster and field are 84 and 164, respectively, whilst the number of galaxies identified by the C.L.U.M.P.S algorithm are 79 and 184. We find no significant difference in the stellar mass and cluster-centric radial distributions of both the field and cluster galaxy samples selected by either of these membership algorithms\footnote{A two-sample KS test does not reject the null hypothesis that the cluster samples produced by the two  membership algorithms are drawn from the same distribution with a $p$-values of $>0.99$ for both stellar mass and cluster-centric radius. The same is also found for the field samples produced by the  membership algorithms.}. We also confirm that all of the conclusions in this work remain regardless of whether the membership is defined using both the Clean and C.L.U.M.P.S algorithms or solely the Clean or C.L.U.M.P.S algorithm.

\section{Comparison to H\texorpdfstring{$\alpha$}{alpha}-derived star formation rates}
\label{sec:appendix_Zeimann_comparison}
To further explore our findings regarding the environmental dependence of the star-forming main sequence in the context of studies at the same epoch derived from SFR proxies other then [\ion{O}{II}] emission, we compare with data from \cite{Zeimann_2013}. In this study, \cite{Zeimann_2013} compare the star-forming main sequence of galaxies across 18 galaxy clusters at 1.0 < z < 1.5. We perform the same procedure as described in Section~\ref{sec:Results}, using the full galaxy GOGREEN sample main sequence relation to calculate $\Delta{\rm SFR}_{\rm MS}$ values for 71 cluster and 71 field galaxies from \cite{Zeimann_2013}. 

\begin{figure}
	\includegraphics[width=1.0\columnwidth]{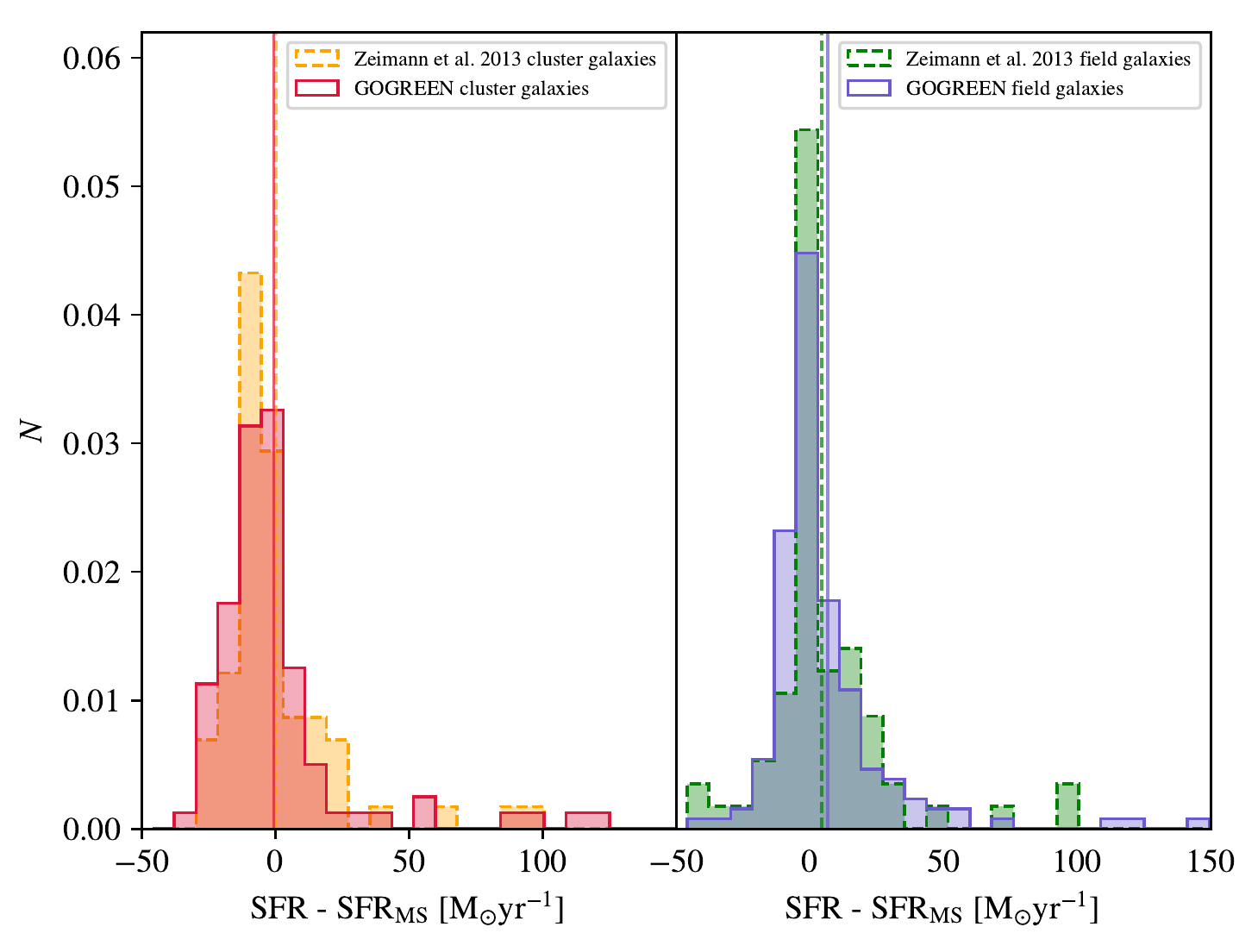}
    \caption{[\ion{O}{II}]-derived GOGREEN cluster (solid) $\rm \Delta SFR_{\rm MS}$ distribution  and H$\alpha$-derived \citet{Zeimann_2013} cluster (dashed) $\rm \Delta SFR_{\rm MS}$ distribution (\emph{left}). GOGREEN field (solid) and \citet{Zeimann_2013} field (dashed) $\rm \Delta SFR_{\rm MS}$ distributions are shown on the \emph{right}.}
    \label{fig:joint_delta_MS_SFR_hist_80pc_flux_lim_flux_cal_SFR_lim_yscale_density_true_vs_Zeimann_2013_field}
\end{figure}

From Figure~\ref{fig:joint_delta_MS_SFR_hist_80pc_flux_lim_flux_cal_SFR_lim_yscale_density_true_vs_Zeimann_2013_field}, we see that the [\ion{O}{II}]-derived GOGREEN and H$\alpha$-derived \citet{Zeimann_2013} $\rm \Delta SFR_{\rm MS}$ distributions (\emph{right}) are remarkably similar in terms of the mean $\rm \Delta SFR_{\rm MS}$. We also directly compare the [\ion{O}{II}]-derived GOGREEN and H$\alpha$-derived \citet{Zeimann_2013} star-forming main sequence, again finding remarkable agreement between the [\ion{O}{II}]-derived and H$\alpha$-derived SFRs. As discussed in \citet{Zeimann_2013}, there is no significant difference between cluster and field star-forming main sequence in the \citet{Zeimann_2013} sample.
\begin{figure}
	\includegraphics[width=1.0\columnwidth]{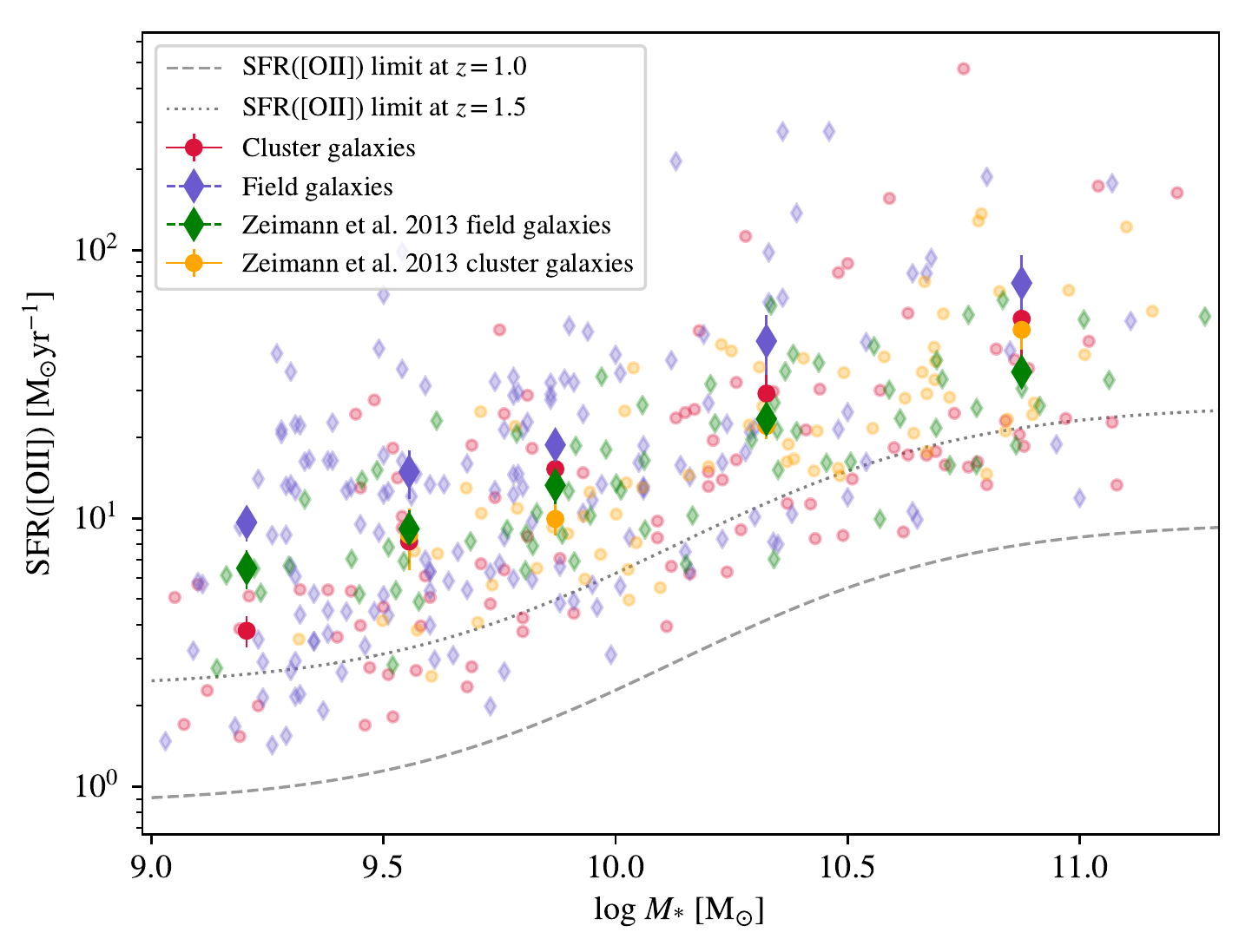}
    \caption{[\ion{O}{II}]-derived main sequence of star formation of cluster galaxies versus field galaxies in the GOGREEN fields, along with the H$\alpha$-derived main sequence from \citet{Zeimann_2013}. The solid purple and crimson markers signify the mean GOGREEN field SFRs and the GOGREEN cluster galaxy SFRs in each stellar mass bin respectively. The solid orange and green markers signify the mean \citet{Zeimann_2013} field SFRs and the \citet{Zeimann_2013} cluster galaxy SFRs in each stellar mass bin respectively. The GOGREEN field SFRs have been corrected using the cosmic SFR vs. $z$ relation of Equation~\ref{eq:Schreiber_2015_SFR} in order to match the mean redshift of GOGREEN cluster galaxies within each stellar mass bin. The error bars represent the bootstrap standard error from bootstrap resampling the data within each bin. The dashed and dotted grey lines represent the SFRs that correspond to the $80\%$ flux completeness limit at $z=1.0$ and $z=1.5$ respectively. }
    \label{fig:log_Mstel_vs_SFR_binned_80pc_flux_lim_flux_cal_z_corr_schreiber15_bootstrap_lines_Zeimann}
\end{figure}

\section{Toy model cluster infall redshift distribution}
\label{sec:appendix_cluster_infall}
In order to generate a parent sample of galaxies whose infall redshifts correspond to the time at which the galaxies were accreted into the cluster, we employ a physically-motivated distribution of infall redshifts, following \citet{Neistein_2006, Neistein_2008}, using a functional form of the average mass accretion history similar to that of the main progenitor (MP) halo in the Extended Press-Schechter formalism \citep{Press_1974, Bond_1991, Lacey_1993} where the growth rate is given by:
\begin{equation}
 \frac{dM_{\rm 12}}{d\omega}=-\alpha M^{1+\beta}_{\rm 12},
  \label{eq:Neistein_2008_Eq7}
\end{equation}
\noindent where the time variable, $\omega\equiv \delta_{\rm c}(z)/D(z)$ with $\delta_{\rm c}(z) \simeq 1.69$ and $D(z)$ is the cosmological linear growth rate. In this parameterisation, $M_{\rm 12}=\langle M_{\rm 1}\rangle/10^{\rm 12} h^{-1}\rm{M_{\odot}}$ where $\langle M_{\rm 1}\rangle$ is the average mass of the MP. The best-fitting parameters derived from the halo statistics in the Millennium Simulation are $\alpha=0.59$ and $\beta=0.141$. The growth rate can be expressed in terms of time via: 
\begin{equation}
\frac{dM_{\rm 12}}{dt}=\dot{\omega}\frac{dM_{\rm 1}}{d\omega},
  \label{eq:Neistein_2008_Eq8}
\end{equation}
where $\dot{\omega}$ is approximated as:
\begin{equation}
 \dot{\omega}=-0.0470[1+z+0.1(1+z)^{\rm 1.25}]^{2.5} h_{\rm 73} \rm Gyr^{\rm -1}.
  \label{eq:Neistein_2008_Eq9}
\end{equation}
\noindent We link the smooth accretion of mass growth over time to the accretion of galaxies by assigning infall redshifts to the model galaxy population in a manner which ensures that the slope of the redshift versus halo mass growth relation matches the slope of the infall redshift distribution of the model galaxy population.


\bsp	
\label{lastpage}
\end{document}